\documentclass[11pt,a4paper]{article}
\pdfoutput=1
\usepackage{jheppub}
\usepackage[T1]{fontenc}
\usepackage{fix-cm}
\usepackage{lmodern}
\usepackage{amsmath,amssymb,mathtools,mathrsfs,empheq}
\usepackage{hyperref}
\usepackage[utf8]{inputenc}
\usepackage{textcomp}
\usepackage[english]{babel}
\usepackage{physics}
\usepackage{tensor}
\usepackage{bbold}
\usepackage{enumerate}
\usepackage{subcaption}
\usepackage{graphicx}
\usepackage{etoolbox}
\usepackage{bm}
\AtBeginEnvironment{pmatrix}{\everymath{\displaystyle}}
\AtBeginEnvironment{bmatrix}{\everymath{\displaystyle}}
\DeclareMathAlphabet{\mathpzc}{OT1}{pzc}{m}{it}
\usepackage{tikz}
\usepackage{xcolor}
\usepackage{comment}

\def\iu{\ensuremath{\mathrm{i}}}

\def\be{\begin{equation}}
\def\ee{\end{equation}}
\def\barray{\begin{array}}
\def\earray{\end{array}}
\def\be{\begin{equation}}
\def\ee{\end{equation}}
\def\bea{\begin{eqnarray}}
\def\eea{\end{eqnarray}}
\def\bal{\begin{align}}
\def\eal{\end{align}}

\usepackage{tensor}

\title{Wrapped M5-branes and AdS$_5$ Black Holes}
\author[a]{Nikolay Bobev,}
\author[a]{Vasil Dimitrov,}
\author[a]{and Annelien Vekemans}

\affiliation[a]{Instituut voor Theoretische Fysica, KU Leuven,\\
Celestijnenlaan 200D, B-3001 Leuven, Belgium}

\emailAdd{nikolay.bobev; vasko.dimitrov; annelien.vekemans@kuleuven.be}

\abstract{We use consistent truncations in supergravity to show that the backreaction of $N$ rotating M5-branes wrapped on a Riemann surface leads to asymptotically AdS$_5$ black hole solutions of 11d supergravity. We discuss the thermodynamic properties of these black holes focusing on their supersymmetric limit. The Bekenstein-Hawking entropy of the supersymmetric black holes scales as $N^3$ and can be reproduced by the superconformal index of the holographically dual 4d $\mathcal{N}=1$ SCFTs of class $\mathcal{S}$.}

\makeatletter
\let\save@mathaccent\mathaccent
\newcommand*\if@single[3]{%
  \setbox0\hbox{${\mathaccent"0362{#1}}^H$}%
  \setbox2\hbox{${\mathaccent"0362{\kern0pt#1}}^H$}%
  \ifdim\ht0=\ht2 #3\else #2\fi
  }
\newcommand*\rel@kern[1]{\kern#1\dimexpr\macc@kerna}
\newcommand*\widebar[1]{\@ifnextchar^{{\wide@bar{#1}{0}}}{\wide@bar{#1}{1}}}
\newcommand*\wide@bar[2]{\if@single{#1}{\wide@bar@{#1}{#2}{1}}{\wide@bar@{#1}{#2}{2}}}
\newcommand*\wide@bar@[3]{%
  \begingroup
  \def\mathaccent##1##2{%
    \let\mathaccent\save@mathaccent
    \if#32 \let\macc@nucleus\first@char \fi
    \setbox\z@\hbox{$\macc@style{\macc@nucleus}_{}$}%
    \setbox\tw@\hbox{$\macc@style{\macc@nucleus}{}_{}$}%
    \dimen@\wd\tw@
    \advance\dimen@-\wd\z@
    \divide\dimen@ 3
    \@tempdima\wd\tw@
    \advance\@tempdima-\scriptspace
    \divide\@tempdima 10
    \advance\dimen@-\@tempdima
    \ifdim\dimen@>\z@ \dimen@0pt\fi
    \rel@kern{0.6}\kern-\dimen@
    \if#31
      \overline{\rel@kern{-0.6}\kern\dimen@\macc@nucleus\rel@kern{0.4}\kern\dimen@}%
      \advance\dimen@0.4\dimexpr\macc@kerna
      \let\final@kern#2%
      \ifdim\dimen@<\z@ \let\final@kern1\fi
      \if\final@kern1 \kern-\dimen@\fi
    \else
      \overline{\rel@kern{-0.6}\kern\dimen@#1}%
    \fi
  }%
  \macc@depth\@ne
  \let\math@bgroup\@empty \let\math@egroup\macc@set@skewchar
  \mathsurround\z@ \frozen@everymath{\mathgroup\macc@group\relax}%
  \macc@set@skewchar\relax
  \let\mathaccentV\macc@nested@a
  \if#31
    \macc@nested@a\relax111{#1}%
  \else
    \def\gobble@till@marker##1\endmarker{}%
    \futurelet\first@char\gobble@till@marker#1\endmarker
    \ifcat\noexpand\first@char A\else
      \def\first@char{}%
    \fi
    \macc@nested@a\relax111{\first@char}%
  \fi
  \endgroup
}
\makeatother

\begin{document}

\maketitle 

\section{Introduction}\label{sec-Introduction}

String and M-theory provide a consistent framework to describe the quantum physics of black holes in terms of microscopic ingredients such as strings and branes. The AdS/CFT correspondence gives an additional vantage point on this venerable research endeavor. Asymptotically AdS black holes can be viewed through the prism of holography and their properties can in principle be delineated using the quantum dynamics of a dual CFT. This is best understood when the AdS black hole solution is constructed in a consistent supergravity theory that arises as a low-energy limit of string or M-theory. This allows to understand the black hole solution in terms of its microscopic ingredients, identify its dual CFT, and then harness the power of holography to study the black hole itself. Supersymmetry can be a very useful crutch in this enterprise. It is often technically easier to construct explicit supersymmetric AdS black hole solutions in supergravity and explore their properties. Moreover, the recent advances in supersymmetric localization allow for the explicit calculation of the path integral of holographic SCFTs, which in turn provides a host of information about the physics of the black hole. Indeed, recently there has been a flurry of activity using these methods to study the thermodynamic and microscopic properties of supersymmetric AdS black holes and their dual QFT description, see \cite{Zaffaroni:2019dhb} and references therein for a review. 

Our goal in this work is to contribute to these developments by studying supersymmetric AdS$_5$ black holes in M-theory that arise from the backreaction of M5-branes wrapped on a compact Riemann surface, $\Sigma_{\mathfrak{g}}$, of genus $\mathfrak{g}$. The supergravity solutions describing this system of branes can be constructed in a two-step procedure that involves supergravity consistent truncations. As shown in \cite{Nastase:1999cb}, reducing 11d supergravity on $S^4$ leads to the maximal 7d $SO(5)$ gauged supergravity of \cite{Pernici:1984xx}. This theory admits an infinite family of supersymmetric AdS$_5$ vacua first explored in the seminal work of Maldacena-N\'u\~nez  \cite{Maldacena:2000mw} and then generalized in \cite{Bah:2011vv,Bah:2012dg}. In addition to the number of M5-branes, $N$, these AdS$_5$ solutions are characterized by the genus of the Riemann surface, $\mathfrak{g}$, and a rational number $z$. We show that for each value of $(\mathfrak{g},z)$ the 7d gauged supergravity theory admits a consistent truncation to 5d $\mathcal{N}=2$ minimal gauged supergravity. This type of consistent truncations were studied previously in the literature, see \cite{Gauntlett:2007sm,Szepietowski:2012tb,MatthewCheung:2019ehr,Cassani:2019vcl,Faedo:2019cvr,Cassani:2020cod,Malek:2020jsa,Josse:2021put}, and our analysis adds several technical details to these constructions including explicit formulae that allow to uplift any solution of the 5d supergravity theory to 11d. 

It is well-known that 5d $\mathcal{N}=2$ minimal gauged supergravity admits supersymmetric asymptotically AdS$_5$ black hole solutions with one electric charge and two independent angular momenta \cite{Gutowski:2004ez,Chong:2005hr}. The sequence of consistent truncations described above provides an embedding of these supersymmetric black holes in M-theory and associates a family of black hole solutions to each of the supersymmetric AdS$_5$ vacua found in \cite{Bah:2011vv,Bah:2012dg}. As emphasized and carefully studied in \cite{Cabo-Bizet:2018ehj,Cassani:2019mms}, the thermodynamic properties of these black holes are somewhat subtle to study since they have finite entropy and on-shell action even in the zero temperature supersymmetric limit. We build on these results and show how the entropy, on-shell action, and charges of the black hole solution can be expressed in terms of the microscopic parameters $(N,\mathfrak{g},z)$ that define the underlying M5-brane system. The black hole solutions admit a somewhat subtle analytic continuation to Euclidean signature which allows for rigorous calculations of their properties using the tools of holographic renormalization which we discuss in some detail. The asymptotic boundary of these Euclidean saddles points of the gravitational path integral has $S^1\times S^3$ topology and points to a description of this system in terms of the superconformal index of a dual 4d $\mathcal{N}=1$ SCFT.

Indeed, we show that these supersymmetric gravitational solutions can be described by the large $N$ limit of the 4d $\mathcal{N}=1$ class $\mathcal{S}$ SCFTs constructed in \cite{Gaiotto:2009gz,Benini:2009mz,Bah:2011vv,Bah:2012dg}. The different values of $z$ that label these theories correspond to the family of topological twists of the 6d $\mathcal{N}=(2,0)$ SCFT living on the worldvolume of the M5-branes that preserve 4d $\mathcal{N}=1$ supersymmetry. The superconformal index of the class $\mathcal{S}$ theories can be studied at low values of $N$ using the results in \cite{Beem:2012yn,Gadde:2009kb,Gadde:2011ik,Rastelli:2014jja}, however we are not aware of a detailed study of the index at large $N$ using these methods. Fortunately we can bypass this impasse. It was realized recently that the superconformal index of general 4d $\mathcal{N}=1$ SCFTs exhibits universal properties that prove very useful in a holographic setup like ours. In particular, it was shown in \cite{Cassani:2021fyv}, see also \cite{GonzalezLezcano:2020yeb}, that carefully treating the superconformal index as a complex function of its fugacities leads to a simple universal formula for it in the so-called Cardy-like limit where the radius of the $S^1$ is much smaller than that of $S^3$. This universal expression for the superconformal index is entirely controlled by the 't Hooft anomalies of the 4d $\mathcal{N}=1$ SCFT and as discussed recently in \cite{Bobev:2022bjm,Cassani:2022lrk} can be reproduced by a similarly universal supergravity on-shell action for large $N$ holographic SCFTs.\footnote{As discussed below, it is not clear to us why the large $N$ and Cardy-like limit have an overlapping regime of validity but the results we present here, as well as the ones in \cite{Bobev:2022bjm,Cassani:2022lrk}, point to a more general regime of validity of the ``second sheet'' formula for the index derived in \cite{Cassani:2021fyv}.} These results have important implications for our wrapped M5-brane setup. We use the 't Hooft anomalies of the 4d $\mathcal{N}=1$ class $\mathcal{S}$ SCFTs derived in \cite{Gaiotto:2009gz,Benini:2009mz,Bah:2011vv,Bah:2012dg} to write the leading $N^3$ term of the superconformal index on ``the second sheet'' in terms of the microscopic parameters $(N,\mathfrak{g},z)$. This expression then exactly agrees with the on-shell action of the supersymmetric CCLP solution computed via holographic renormalization. An appropriate Legendre transformation of this index then yields an expression for the microscopic entropy of the dual black hole which perfectly agrees with the Bekenstein-Hawking entropy formula. This constitutes a precision test of holography and a microscopic account of the wrapped M5-brane black hole microstates in terms of the dual class $\mathcal{S}$ SCFT.

The rest of this paper is organized as follows. In Section~\ref{sec-M-to-7} we summarize some results on the dimensional reduction and consistent truncation from 11d supergravity to 7d gauged supergravity. In Section~\ref{sec-truncation} we derive a family of dimensional reductions on $\Sigma_{\mathfrak{g}}$ from this 7d theory to 5d $\mathcal{N}=2$ minimal gauged supergravity. Further details on the reduction of the supergravity action and BPS equations are given in Appendix~\ref{appendix7d5d}. In Section~\ref{sec-thermo-CCLP}, supplemented by Appendix \ref{sec-CCLP-appendix}, we provide some details on mostly known results regarding the thermodynamics of the most general known rotating black hole solution of the 5d gauged supergravity theory and the calculation of its on-shell action. Section~\ref{sec-holography} is devoted to the holographic interpretation of these results and their connection to the superconformal index of the 4d $\mathcal{N}=1$ class $\mathcal{S}$ SCFTs. In Section~\ref{sec-Discussion} we provide some concluding comments, while in Appendix~\ref{sec-conventions} we summarize our conventions.


\section{7d gauged supergravity from 11d}
\label{sec-M-to-7}

The bosonic sector of the 11d supergravity contains the metric with line element $\dd{s}^2_{11}$ and a three-form $\mathcal{A}$. The bosonic action of the theory is given by
\begin{equation}\label{11DSUGRAaction}
    I_{11} = \frac{1}{16 \pi G_{11}}\int \qty(  \star_{11} R - \frac{1}{2} \mathcal{F} \wedge \star_{11} \mathcal{F} + \frac{1}{6}\mathcal{F} \wedge \mathcal{F} \wedge \mathcal{A} ) \,,
\end{equation}
where $ \mathcal{F} = \dd{\mathcal{A}}$. As shown in \cite{Nastase:1999cb}, see also \cite{Donos:2010ax} for some more details, a reduction of 11d supergravity on $ S^4$ results in a consistent truncation to the 7d, $ \mathcal{N}=4$, $ SO(5)$ gauged supergravity of \cite{Pernici:1984xx}. The 7d gauged supergravity can be further consistently truncated to its $ U(1) \times U(1)$ invariant sector as described in \cite{Liu:1999ai}. The bosonic field content of this 7d theory, which plays an important role below, consists of the metric with line element $\dd{s}^2_7$, together with a three-form, two Abelian one-forms and two scalars: $ \qty(S, \, A_1, \, A_2, \, \lambda_1, \, \lambda_2)$. 

We now provide some more details on how this sequence of consistent truncations works. We parametrize the $S^4$ by the four angles: $ (\alpha, \, \beta, \, \phi_1, \, \phi_2)$, as
\begin{align}\label{eq:S4metric}
    \dd{s}^2_4 & = \frac{1}{\mu^2} \qty( e^{-2\lambda_1} \qty[ \dd{\nu_1}^2 + \nu_1^2 (D \phi_1)^2 ] + e^{-2\lambda_2} \qty[ \dd{\nu_2}^2 + \nu_2^2 (D \phi_2)^2 ] + e^{4\lambda_1 + 4\lambda_2} \dd{\nu_3}^2) \,,
\end{align}
where $ \mu$ is an inverse length scale, $ \nu_{1,2,3}$ are alternative (constrained) coordinates for the two angles $ \qty(\alpha,\beta)$:
\begin{align}
    \nu_1 & = \sin \alpha \cos \beta \,, \quad \nu_2 = \sin \alpha \sin \beta \,, \quad \nu_3 = \cos \alpha \,;  \qquad \nu_1^2 + \nu_2^3 + \nu_3^2 = 1 \,,
\end{align}
and $ D\phi_{1,2} = \dd{\phi_{1,2}} + \mu A_{1,2}$. The coordinates $ \phi_{1,2}$ parametrize the $ U(1) \times U(1)$ isometry directions. The 11d fields are expressed in terms of the 7d fields as
\begin{align}\label{eq-11d-final-metric-and-G}
    \dd{s}^2_{11} & = \Delta^{1/3} \dd{s}^2_7 + \Delta^{-2/3} \dd{s}^2_4 \,, \nonumber \\
    \mathcal{F} & = - e^{-4\lambda_1-4\lambda_2} \nu_3 \star_7 S + \frac{1}{\mu} S \wedge \dd{\nu_3} \nonumber \\ 
    & \quad + \frac{1}{\Delta \mu^2} \Bigg[ - e^{2\lambda_1} \nu_1^2 \, D \phi_1 \wedge F_2 \wedge \dd{\nu_{3}} - e^{2\lambda_2} \nu_2^2 \, D \phi_2 \wedge F_1 \wedge \dd{\nu_{3}} \nonumber \\
    & \qquad \qquad \qquad \, + e^{-4\lambda_1-4\lambda_2} \nu_1 \nu_3 \, D \phi_1 \wedge F_2 \wedge \dd{\nu_1} + e^{-4\lambda_1-4\lambda_2} \nu_2 \nu_3 \, D \phi_2 \wedge F_1 \wedge \dd{\nu_2} \Bigg] \nonumber \\
    & \quad + \frac{U \nu_1 \nu_2}{\Delta^2\mu^3} \Bigg[ \nu_1 \, D \phi_1 \wedge D \phi_{2} \wedge \dd{\nu_2} \wedge \dd{\nu_{3}} + \nu_2 \, D \phi_2 \wedge D \phi_1 \wedge \dd{\nu_1} \wedge \dd{\nu_{3}} \nonumber \\
    & \qquad \qquad \qquad + \nu_3 \, D \phi_1 \wedge D \phi_{2} \wedge \dd{\nu_1} \wedge  \dd{\nu_2} \Bigg]  \nonumber\\
    & \quad + \frac{\nu_1\nu_2}{\Delta^2\mu^3} \Bigg[ 2 e^{2\lambda_1 + 2\lambda_2} \nu_1 \nu_2 \, D \phi_1 \wedge D \phi_2 \wedge \dd{\qty(\lambda_1 - \lambda_2)} \wedge \dd{\nu_3}  \nonumber \\
    & \qquad \qquad \qquad + 2 e^{-2\lambda_1 - 4\lambda_2} \nu_1 \nu_3 \, D \phi_1 \wedge D \phi_2 \wedge \dd{(3\lambda_1 - 2\lambda_2)} \wedge \dd{\nu_2} \nonumber\\
    & \qquad \qquad \qquad + 2 e^{-4\lambda_1 - 2\lambda_2} \nu_2 \nu_3 \, D \phi_1 \wedge D \phi_2 \wedge \dd{(3\lambda_1 + 2\lambda_2)} \wedge \dd{\nu_1} \Bigg] \,,
\end{align}
where $ F_{1,2} = \dd{A_{1,2}}$ and $ \qty(\Delta, U)$ are functions given by
\begin{align}
    \begin{aligned}
        \Delta & = e^{2\lambda_1} \nu_1^2 + e^{2\lambda_2} \nu_2^2 + e^{-4\lambda_1 - 4 \lambda_2} \nu_3^2 \,, \\
        U & = -\qty(2e^{2\lambda_1 + 2\lambda_2} + e^{-2\lambda_1 - 4\lambda_2}) \nu_1^2 -(2e^{2\lambda_1 + 2\lambda_2} + e^{-4\lambda_1 - 2\lambda_2}) \nu_2^2  \\
        & \qquad + \qty(e^{-8\lambda_1 - 8\lambda_2} - 2e^{-2\lambda_1 - 4\lambda_2} - 2 e^{-4\lambda_1 - 2\lambda_2}) \nu_3^2 \,.
    \end{aligned}
\end{align}
The resulting 7d theory has the following bosonic action
\begin{align}\label{7DSUGRAaction}
    I_{7} & = \frac{1}{16\pi G_7}\int \Bigg[  \star_7 (R - V) - 5 \dd{\lambda_+}\wedge \star_7 \dd{\lambda_+} - \dd{\lambda_-}\wedge \star_7 \dd{\lambda_-} - \frac{1}{2} e^{-4\lambda_1}F_1\wedge \star_7 F_1 \nonumber\\
    & \qquad \qquad \qquad \quad - \frac{1}{2} e^{-4\lambda_2}F_2\wedge \star_7 F_2 - \frac{1}{2} e^{-4\lambda_+} S\wedge \star_7 S + \frac{1}{2\mu} \, S\wedge \dd{S} \nonumber\\
    & \qquad \qquad \qquad \quad - \frac{1}{\mu} F_1\wedge F_2 \wedge S + \frac{1}{4\mu}\qty(A_1\wedge F_1\wedge F_2\wedge F_2 + A_2 \wedge F_2 \wedge F_1 \wedge F_1) \Bigg],
\end{align}
where we have defined $ \lambda_\pm = \lambda_1 \pm \lambda_2$ and $ V$ is the potential for the scalars, given by
\begin{equation}\label{eq-scalar-potential}
    V = \frac{\mu^2}{2} \qty(e^{-8 \lambda _1-8 \lambda _2}-4 e^{-2 \lambda _1-4 \lambda _2}-4 e^{-4 \lambda _1-2 \lambda _2}-8 e^{2 \lambda _1+2 \lambda _2}) \,.
\end{equation}
Out of this action we obtain the following equations of motion
\begin{align}
    \begin{aligned}
        \dd{S} & = \mu \, e^{-4\lambda_+} S + F_1 \wedge F_2 \,, \\
        \dd{\qty(e^{-4\lambda_{1,2}}  \star_7 F_{1,2})} & = \frac{1}{\mu} F_{1,2} \wedge F_{2,1} \wedge F_{2,1} - \frac{1}{\mu} F_{2,1} \wedge \dd{S} \,, \\
        \dd{ \star_7 \dd{\qty(3 \lambda_{1,2} + 2 \lambda_{2,1})} } & = - \frac{1}{2} e^{-4\lambda_{1,2}} F_{1,2} \wedge  \star_{7} F_{1,2} - \frac{1}{2} e^{-4\lambda_+} S \wedge  \star_7 S + \frac{1}{4} \star_7 \pdv{V}{\lambda_{1,2}} \,, \\
        R_{MN} & = 5 \partial_M \lambda_+ \partial_N \lambda_- + \partial_M \lambda_- \partial_N \lambda_+ + \frac{1}{2} e^{-4\lambda_1} F_{1 \, MN}^{2} + \frac{1}{2} e^{-4\lambda_2} F_{2 \, MN}^{2} \\
        + \frac{1}{4} & e^{-4 \lambda_+} S^2_{MN} + \frac{1}{5}g_{MN}\qty[V - \frac{1}{4} e^{-4\lambda_1} F_1^2 - \frac{1}{4} e^{-4\lambda_2} F_2^2 - \frac{1}{6} e^{-4\lambda_+} S^2] \,,
    \end{aligned}
\end{align}
where $ M,N,\dots$ are curved 7d indices and we have defined $ F_{1,2 \, MN}^2 = F\indices{_{1,2 \, M}^R} F_{1,2 \, N R}, \, S_{MN}^2 = S\indices{_M^{RS}}S_{NRS}, \, F_{1,2}^2 = F_{1,2}^{MN}F_{1,2 \, MN}, \, S^2 = S^{MNR}S_{MNR}$. To ensure that a solution of the equations of motion of the $ U(1) \times U(1)$ invariant truncation of the 7d supergravity theory is supersymmetric one should check that the supersymmetry variations of one spin-$ \frac{3}{2}$ gravitino and two spin-$ \frac{1}{2}$  dilatini, $ \qty(\widetilde{\psi}_M, \, \widetilde{\zeta}_1, \, \widetilde{\zeta}_2)$, vanish. Setting the supersymmetry variations of these fields to zero gives the following BPS equations for the $ 32$-dimensional 7D spinor $ \widetilde{\epsilon} \otimes \mathpzc{E}$, which we have split into a purely 7D part $ \widetilde{\epsilon}$ and an $ SO(5)$ part $ \mathpzc{E}$:
\begin{align}\label{eq-7D-bps}
    \delta \widetilde{\psi}_M & = \Bigg[ \partial_M + \frac{1}{4} \omega_{M \widehat{A} \widehat{B}} \widetilde{\gamma}^{\widehat{A} \widehat{B}} \otimes 1_{4} + \frac{\mu}{2} \, 1_{8} \otimes \qty(A_{1 \, M} \Gamma^{12} + A_{2 \, M} \Gamma^{34}) \nonumber \\
    & \qquad + \qty( \frac{\mu}{4} e^{-4\lambda_+} \widetilde{\gamma}_M + \frac{1}{2} \widetilde{\gamma}_M \widetilde{\gamma}^N \partial_N \lambda_+ ) \otimes 1_{4} \nonumber \\
    & \qquad + \frac{1}{4} \qty(e^{-2\lambda_1} F_{1 \, MN} \widetilde{\gamma}^N \otimes \Gamma^{12} + e^{-2\lambda_2} F_{2 \, MN} \widetilde{\gamma}^N \otimes \Gamma^{34}) \nonumber\\
    & \qquad - \frac{1}{8}  e^{-2\lambda_+} S_{MNP} \widetilde{\gamma}^{NP} \otimes \Gamma^5 \Bigg] \, \widetilde{\epsilon} \otimes \mathpzc{E} \,, \nonumber \\
    \delta\widetilde{\zeta}_{1,2} & = \Bigg[\frac{\mu}{4} \qty(e^{2\lambda_{1,2}} - e^{-4\lambda_+}) - \frac{1}{4} \widetilde{\gamma}^M \partial_M \qty(3\lambda_{1,2} + 2\lambda_{2,1}) \otimes 1_{4} \nonumber \\
    &  \qquad - \frac{1}{16} e^{-2\lambda_{1,2}} F_{1,2 \, MN} \widetilde{\gamma}^{MN} \otimes \Gamma^{12} + \frac{1}{48} e^{-2\lambda_+}  S_{MNP} \widetilde{\gamma}^{MNP} \otimes \Gamma^5\Bigg] \, \widetilde{\epsilon} \otimes \mathpzc{E} \,.
\end{align}
Here $ \widetilde{\gamma}^M$ is an $ 8 \times 8$ curved 7d spacetime gamma matrix, $ \widetilde{\gamma}^{\widehat{A}}$ is a flat 7d Lorentz gamma matrix, $ \omega_{M \widehat{A} \widehat{B}}$ is the 7d spin connection, $ \Gamma^i$, with $ i = 1,\dots,5$, is a $ 4 \times 4$ flat internal $ SO(5)$ gamma matrix, and finally $ 1_{n}$ is an $ n \times n$ identity matrix.

\section{Dimensional reduction to 5d $ \mathcal{N} = 2$ minimal gauged supergravity}
\label{sec-truncation}

As shown in \cite{Maldacena:2000mw,Bah:2011vv,Bah:2012dg} the 7d $U(1) \times U(1)$ invariant truncation discussed above admits an infinite family of supersymmetric AdS$_5$ vacua. This suggests that there is a further consistent truncation of this model to 5d $ \mathcal{N} = 2$ minimal gauged supergravity. Indeed, it is possible to construct such a consistent truncation for each of the supersymmetric vacua studied in \cite{Bah:2011vv,Bah:2012dg}. This type of consistent truncations arising from wrapped M5-branes have been discussed in various level of detail in \cite{Gauntlett:2007sm,Szepietowski:2012tb,MatthewCheung:2019ehr,Cassani:2019vcl,Faedo:2019cvr,Cassani:2020cod,Malek:2020jsa,Josse:2021put}. Our goal here is to rederive some of these results in a way tailored to our goal of discussing asymptotically AdS$_5$ black hole solutions and where needed fill in small gaps in the literature.

More concretely, our aim is to reduce the 7d $ U(1) \times U(1)$ gauged supergravity truncation on a smooth Riemann surface $\Sigma_{\mathfrak{g}}$ with genus $ \mathfrak{g}$ to arrive at 5d $ \mathcal{N}=2$ minimal gauged supergravity, with bosonic field content given by the metric with line element $ \dd{s}^2_5$ and a single one-form gauge field $ A$. For concreteness, we take the Riemann surface to be hyperbolic, i.e. $ \mathfrak{g} > 1$,\footnote{Our results can be adapted to the AdS$_5$ vacua of \cite{Bah:2011vv,Bah:2012dg} with $ \mathfrak{g} =0,1$ in a straightforward way.} and parametrize it with coordinates $ (x_1, \, x_2)$ as:\footnote{To have a compact Riemann surface we have to mod out the upper half plane with an appropriate discrete group. From now on we will assume that this discrete identification has been performed and will work with local coordinates on the Riemann surface.}
\begin{align}\label{eq:H2met}
        \dd{s}^2_2 & = \qty(\dd{x_1}^2 + \dd{x_2}^2)/x_2^2 \,.
\end{align}
We also define the one-form $ \omega = \dd{x_1}/x_2$, such that the volume form on $ \Sigma_{\mathfrak{g}}$ is given by $ \dd{\omega}$. We take the 7d scalars to assume constant values $ \lambda_{1,2} = \Lambda_{1,2}$ as dictated by the AdS$_5$ vacua of \cite{Bah:2011vv,Bah:2012dg} and define two new constants, $ (g_0, \, f_0)$, in terms of them
\begin{align}\label{eq-constants-ansatz}
    \begin{aligned}
    \Lambda_1 & = \frac{1}{10} \ln \qty(\frac{1+ 7z + 7z^2 +33 z^3 - (1+4 z + 19 z^2) \sqrt{1+ 3z^2}}{4z(1-z)^2}) \,, \\
    \Lambda_2 & = \Lambda_1 - \frac{1}{2} \ln \qty(\frac{1+z}{2z + \sqrt{1+3z^2}}) \,, \\
    f_0 & = 4\Lambda_+ \,, \\
    g_0 & = \frac{1}{2} \ln\qty(\frac{1}{8} e^{2\Lambda_+} \qty[(1-z)e^{2\Lambda_1} + (1+z) e^{2\Lambda_2}]) \,,
    \end{aligned}
\end{align}
where $ \Lambda_\pm = \Lambda_1 \pm \Lambda_2$ and $ z$ is a rational number obeying $z(\mathfrak{g}-1) \in \mathbb{Z}$. In this way we will obtain a family of dimensional reductions labelled by $ \qty(z, \mathfrak{g})$. For the remaining 7d fields we take the following reduction Ansatz
\begin{align}\label{eq-ansatz}
    \begin{aligned}
        \dd{s}^2_7 & = e^{2f_0} \, \dd{s}^2_5 + e^{2g_0} L^2 \, \dd{s}^2_2 \,, \\
        S & = \mathpzc{c} \, e^{f_0 + 2g_0 + 2 \Lambda_+} L^2 \, \star_7 \qty(F \wedge \dd{\omega}) \,, \\
        A_{1,2} & = \mathpzc{a} \, (1 \pm z)L \, \omega + \mathpzc{b} \,  e^{f_0 + 2\Lambda_{1,2}} \, A \,,
    \end{aligned}
\end{align}
where $ F = \dd{A}$, $ L = 2/\mu$ is the $ \text{AdS}_{5}$ radius, and $ (\mathpzc{a}, \, \mathpzc{b}, \, \mathpzc{c})$ are $ z$-independent constants that we will fix below. We present the details of the dimensional reduction based on this Ansatz in Appendix~\ref{proofappendix}. The upshot is that to have a 5d theory with $ z$-independent cosmological constant one has to fix $ \mathpzc{a} = \pm 1/4$. To retain the normalization of the 5d Maxwell field such that $ F = \dd{A}$ we need to take $ \mathpzc{c} = \mathpzc{b}$ and we also found it convenient to rescale $\mathpzc{b}$ as $ \mathpzc{b} = \mathpzc{x}/\sqrt{3}$. As shown in Appendix~\ref{proofappendix}, integrating over the Riemann surface we obtain the bosonic action of 5d $ \mathcal{N} = 2$ minimal gauged supergravity 
\begin{equation}\label{5DSUGRAaction}
    I_5 = \frac{1}{16\pi G_5}\int\left[ \star_5 \left( R + \frac{12}{L^2} \right) - \frac{\mathpzc{x}^2}{2} \, F \wedge  \star_5 F + \frac{\mathpzc{x}^3}{3\sqrt{3}}A\wedge F\wedge F \right] \,,
\end{equation}
where $ \mathpzc{x}$ can be viewed as a normalization constant associated with the freedom\footnote{Some common choices include: \cite{Chong:2005hr,Cassani:2015upa}: $\mathpzc{x} = 1$, \cite{Gutowski:2004ez}: $\mathpzc{x} = -2$, \cite{Cabo-Bizet:2018ehj}: $\mathpzc{x} = 2/(\sqrt{3}\, g)$.} to rescale the gauge filed $ A$. Importantly, the 5d Newton constant is given in terms of the 7d one by
\begin{equation}\label{eq-G7-G5-relation}
    \boxed{ G_7 = e^{3f_0 + 2g_0}L^2 \, 4\pi(\mathfrak{g}-1) \, G_5 } \,.
\end{equation}
The equations of motion of the 5d supergravity can be derived from the action above and read 
\begin{align}
    \begin{aligned}
        \dd{ \star_5 F} & = \frac{\mathpzc{x}}{\sqrt{3}} F \wedge F \,, \\
        R_{\mu\nu} & = - \frac{4}{L^2} g_{\mu\nu} + \frac{\mathpzc{x}^2}{2}F\indices{_\mu^\alpha}F_{\nu\alpha} - \frac{\mathpzc{x}^2}{12}g_{\mu\nu}F^{\alpha\beta}F_{\alpha\beta} \,,
    \end{aligned}
\end{align}
where we have denoted the 5d curved indices with Greek letters. The only fermionic field in the 5d supergravity theory is the spin-$ \frac{3}{2}$ gravitino $ \psi_\mu$. Setting its supersymmetry variation to zero we obtain the following BPS equation for the $ 4$-dimensional 5d spinor $ \epsilon$: 
\begin{align}\label{eq-bps}
    \delta \psi_\mu & = \qty[\partial_\mu + \frac{1}{4} \omega_{\mu \widehat{\alpha} \widehat{\beta}} \gamma^{\widehat{\alpha} \widehat{\beta}} - \frac{\iu \, \mathpzc{x}}{8 \sqrt{3}}\qty(\gamma\indices{_\mu_{\alpha\beta}}F^{\alpha\beta} - 4F_{\mu\alpha} \gamma^\alpha) - \frac{1}{2L}\qty(\gamma_\mu + \iu \, \sqrt{3}\, \mathpzc{x} A_\mu )] \epsilon \,,
\end{align}
where $ \gamma^\mu$ is a $ 4 \times 4$ curved 5d spacetime gamma matrix, $ \gamma^{\widehat{\alpha}}$ is a flat 5d Lorentz gamma matrix and $ \omega_{\mu \widehat{\alpha} \widehat{\beta}}$ is the 5d spin connection. In Appendix~\ref{sec-reducing-bps} we show that the three 7d supersymmetry variations in (\ref{eq-7D-bps}) reduce to (\ref{eq-bps}) provided we identify
\begin{align}
    \widetilde{\gamma}_{\mu} & = - \chi_7 \, \gamma_{\mu} \otimes \sigma_1 \,, \quad \widetilde{\gamma}_{\widehat{x}_1} = \chi_7 \, 1_4 \otimes \sigma_2 \,, \quad \widetilde{\gamma}_{\widehat{x}_2} = \chi_7 \, 1_4 \otimes \sigma_3 \,, \quad \widetilde{\epsilon} = \epsilon  \otimes \mqty(
    1 \\ 
    \chi_7
    ) \,,
\end{align}
which in turn also implies that $ \widetilde{\gamma}_{\widehat{x}_1 \widehat{x}_2} = \iu \, 1_4 \otimes \sigma_1$, use the projectors
\begin{align}
    \qty(\widetilde{\gamma}_{\widehat{x}_1 \widehat{x}_2} \otimes \Gamma^{12}) (\widetilde{\epsilon} \otimes \mathpzc{E})  = \chi_7 \, \widetilde{\epsilon} \otimes \mathpzc{E} \,, \qquad \qty(1_{8} \otimes \Gamma^5) \qty(\widetilde{\epsilon} \otimes \mathpzc{E}) = \widetilde{\epsilon} \otimes \mathpzc{E} \,,
\end{align}
and set
\begin{align}
    \mathpzc{a} & = - \frac{\chi_7}{4} \,, \quad \mathpzc{c} = \mathpzc{b} = \frac{\mathpzc{x}}{\sqrt{3}} \,, \quad \partial_{x_1} \qty(\widetilde{\epsilon} \otimes \mathpzc{E}) = \partial_{x_2} \qty(\widetilde{\epsilon} \otimes \mathpzc{E}) = 0 \,,
\end{align}
where $ \chi_7 = \pm 1$ is the chirality of the highest rank 7d Clifford algebra element. Note that the choice of sign leftover in $ \mathpzc{a}$ from reducing the action is dictating which of the two inequivalent representation of the 7d Clifford algebra one should use.

\section{Thermodynamics of the 5d black hole}\label{sec-thermo-CCLP}

After showing that to each of the supersymmetric AdS$_5$ vacua constructed in \cite{Bah:2011vv,Bah:2012dg} one can associate a distinct truncation of 11d supergravity to 5d $\mathcal{N}=2$ minimal gauged supergravity we now proceed to discuss a family of rotating asymptotically AdS$_5$ black hole solutions of this theory. Although our discussion will be in five dimensions we stress that these solutions can be explicitly uplifted to 11d using the formulae in Section~\ref{sec-M-to-7} and Section~\ref{sec-truncation} above. Many of the results presented below have appeared before in the literature, in particular in \cite{Chong:2005hr,Papadimitriou:2005ii,Cabo-Bizet:2018ehj,Cassani:2019mms}. Our goal is to collect them here for completeness and clarify some aspects of the calculations.\footnote{Most of the supergravity, holographic renormalization, and black hole thermodynamics calculations were performed with the help of a new \texttt{Mathematica} package we developed which is based on \texttt{xAct} and can be found at \url{https://github.com/waskou/SolutionsX}.}

\subsection{CCLP solution}

The 5d $\mathcal{N} = 2$ gauged supergravity with bosonic action (\ref{5DSUGRAaction}) admits a general non-supersymmetric black hole solution found in \cite{Chong:2005hr} which we will refer to as CCLP. The solution is specified by four real parameters $ (a,b,q,m)$ which determine its two angular momentum, electric charge and mass. For the Lorentzian black hole solutions these parameters are further constrained as
\begin{align}\label{eq-param-constraint}
    -1 & < ag < 1 \,, \quad -1 < bg < 1 \,, \quad 0 \leq q \leq \frac{m}{1 + ag + bg} \,,
\end{align}
where $ g = 1/L$ is the inverse of the AdS scale. The first two conditions arise from requiring that the black hole is not over-rotating and the last condition constitutes the BPS bound. The solution is specified by the following metric and gauge field
\begin{align}\label{eq-cclp-hopf1}
    \begin{aligned}
        \dd{s}^2_{\text{CCLP}} & = - \frac{\Delta_\eta [(1 + g^2r^2)\rho^2 \dd{t} + 2q \nu] \dd{t}}{\Xi_a \Xi_b \rho^2} + \frac{2q \nu \omega}{\rho^2} + \frac{f}{\rho^4}\qty(\frac{\Delta_\eta \dd{t}}{\Xi_a \Xi_b} - \omega)^2 \\
        &  \qquad + \frac{\rho^2 \dd{r}^2}{\Delta_r} + \frac{\rho^2 \dd{\eta}^2}{\Delta_\eta} + \frac{r^2 + a^2}{\Xi_a} \sin^2\eta \dd{\xi_1}^2 + \frac{r^2 + b^2}{\Xi_b} \cos^2\eta \dd{\xi_2}^2 \,, \\
        A_{\text{CCLP}} & = \frac{\sqrt{3}\,q}{\mathpzc{x}\rho^2}\qty(\frac{\Delta_\eta \dd{t}}{\Xi_a \Xi_b} - \omega) + \alpha \dd{t} \,,
    \end{aligned}
\end{align}
where we have defined the one-forms 
\begin{align}
    \nu & = b\, \sin^2\eta \dd{\xi_1} + a\, \cos^2\eta \dd{\xi_2} \,, \quad \omega = \frac{a\, \sin^2\eta}{\Xi_a} \dd{\xi_1} + \frac{b\, \cos^2\eta}{\Xi_b} \dd{\xi_2} \,,
\end{align}
the functions
\begin{align}
    \begin{aligned}
        \Delta_r(r) & = \frac{(r^2 + a^2)(r^2 + b^2)(1 + g^2r^2) + q^2 + 2abq}{r^2} - 2m \,, \\
        \Delta_\eta(\eta) & = 1 - a^2g^2\cos^2\eta - b^2g^2\sin^2\eta \,, \\
        \rho^2(r,\eta) & = r^2 + a^2\cos^2\eta + b^2\sin^2\eta \,, \\
        f(r,\eta) & = 2\qty(m + abqg^2) \rho^2(r,\eta) - q^2 \,,
    \end{aligned}
\end{align}
and the constants
\begin{align}
    \Xi_a & = 1 - a^2g^2 \,, \quad \Xi_b = 1 - b^2g^2 \,.
\end{align}
We have also included a pure gauge term, $ \alpha \dd{t}$, in $ A_{\text{CCLP}}$, which is important to guarantee regularity of $ A^\mu A_\mu$ on the horizon. These coordinates cover the region outside of the black hole and have the following ranges and identifications
\begin{align}
    -\infty < t < \infty \,, \quad r_+ < r < \infty \,, \quad 0 < \eta < \frac{\pi}{2} \,, \quad \xi_1 \sim \xi_1 + 2\pi \,, \quad \xi_2 \sim \xi_2 + 2\pi \,,
\end{align}
where $ r_+$ is the largest positive root of $ \Delta_r(r)$. The surface $ r = r_+$ is the event horizon of the Lorentzian black hole, see Appendix \ref{sec-potentials} for further details. Sometimes, it will be convenient to trade the parameter $ m$ for $ r_+$. The explicit relation, which is obtained by solving $ \Delta_r(r_+) = 0$, reads
\begin{align}\label{eq-mTrh}
    m & = \frac{(r_+^2 + a^2)(r_+^2 + b^2)(1 + g^2r_+^2) + q^2 + 2abq}{2r_+^2} \,.
\end{align}

As emphasized in \cite{Cabo-Bizet:2018ehj} the Lorentzian CCLP black hole can be analytically continued to a Euclidean ``black saddle'', which solves the equations of motion of the Euclidean theory (\ref{eq-eom-eucl})\footnote{The term ``black saddle'' was coined in \cite{Bobev:2020pjk} and refers to the fact that, in general, there exist Euclidean saddle points of the supergravity path integral that do not admit analytic continuation to sensible Lorentzian field configurations. The CCLP black saddle is an example of an Euclidean solution that does admit continuation to the Lorentzian black hole of \cite{Chong:2005hr} for some values of the parameters. We refer to this black saddle also as CCLP and the meaning is to be inferred from the context.}.  The Euclidean metric and gauge field are given by
\begin{align}\label{eq-eucl-met-and-A-cclp}
    \dd{s}^2_{\text{CCLP},\mathpzc{E}} = \eval{\dd{s}^2_{\text{CCLP}}}_{t \to -\iu\, \tau} \,, \quad A_{\text{CCLP},\mathpzc{E}} = \eval{A_{\text{CCLP}}}_{t \to -\iu\, \tau} \,.
\end{align}
Note that these metric and gauge field are both complex. One might want to work with a real Euclidean section by, in addition to Wick rotation, also analytically continuing the rotation parameters as: $ a \to \iu\,a, b \to \iu\,b$. Once this is done the Euclidean metric is purely real and the Euclidean gauge field is purely imaginary. In the context of evaluating the on-shell action of this solution it is useful to work with this real Euclidean section. To interpret the result holographically as a Lorentzian partition function the rotation parameters have to be analytically continued back to their original values: $ a \to -\iu\,a, b \to -\iu\,b$. It turns out that this procedure gives the same final answer as simply working with the complex Wick rotated metric and gauge field (\ref{eq-eucl-met-and-A-cclp}), without continuing the rotation parameters back and forth.

In the holographic context it will be useful also to work with the conformally rescaled boundary metric 
\begin{align}\label{eq-conf-bdy}
    \dd{s}^2_{\text{CCLP},\mathpzc{B}} & = \lim_{r \to \infty} g^{-2} r^{-2} \dd{s}^2_{\text{CCLP}} \nonumber\\
    & = - \frac{\Delta_\eta}{\Xi_a \Xi_b} \dd{t}^2 + \frac{1}{g^2 \Delta_\eta} \dd{\eta}^2 + \frac{\sin^2\eta}{g^2 \Xi_a} \dd{\xi_1}^2 + \frac{\cos^2\eta}{g^2 \Xi_b} \dd{\xi_2}^2 \,.
\end{align}
This metric, with certain identifications of the coordinates that are discussed in detail in \cite{Cabo-Bizet:2018ehj}, determines the background on which the 4d $\mathcal{N}=1$ dual SCFT is placed.

\subsection{Thermodynamic properties CCLP}

In Appendix \ref{sec-potentials} we show that to the CCLP black hole we can associate the following thermodynamic potentials:
\begin{align}\label{eq-potentials}
    \begin{aligned}
        \beta & = \frac{2\pi r_+\qty[(r_+^2 + a^2)(r_+^2 + b^2) + abq]}{r_+^4\qty[1 + g^2(2r_+^2 + a^2 + b^2)] - (ab + q)^2} \,, \\
        \Omega_1 & = \frac{a(r_+^2 + b^2)(1 + g^2r_+^2) + bq}{(r_+^2 + a^2)(r_+^2 + b^2) + abq} \,, \quad \Omega_2 = \frac{b(r_+^2 + a^2)(1 + g^2r_+^2) + aq}{(r_+^2 + a^2)(r_+^2 + b^2) + abq} \,, \\
        \Phi & = \frac{\sqrt{3} q r_+^2}{\mathpzc{x}\qty[(r_+^2 + a^2)(r_+^2 + b^2) + abq]} \,,
    \end{aligned}
\end{align}
physically corresponding to: inverse temperature, angular velocities measured in a non-rotating frame at infinity, and electrostatic potential. To obtain the entropy, via the Bekenstein-Hawking area formula $ S = \text{Area}/4G_5$, one has to integrate the square root determinant, $ \sqrt{g_{\mathcal{H}}}$, of the induced metric on the horizon, which has topology of $ S^3$ and metric given by
\begin{align}
    \dd{s}^2_{\mathcal{H}} & = \eval{\dd{s}^2_{\text{CCLP}}}_{\dd{t} \to 0, \,  \dd{r} \to 0, \, r \to r_+} \,.
\end{align}
Performing this integral we obtain
\begin{align}\label{eq-entropy}
    S & = \frac{\pi^2\qty[(r_+^2 + a^2)(r_+^2 + b^2) + abq]}{2\Xi_a \Xi_b r_+ G_5} \,.
\end{align}
In Appendices \ref{sec-charges1} and \ref{sec-charges2} we calculate the following charges of the CCLP black hole
\begin{align}\label{eq-charges}
    \begin{aligned}
        E & = \qty(E_{\text{AdS}} - \widetilde{\zeta}') + \frac{m \pi(2\Xi_a + 2\Xi_b - \Xi_a \Xi_b) + 2abq^2 \pi(\Xi_a + \Xi_b)}{4 \Xi_a^2 \Xi_b^2 G_5} \,, \\
        J_1 & = \frac{\pi\qty[2am + bq(1 + a^2g^2)]}{4\Xi_a^2 \Xi_b G_5} \,, \quad J_2 = \frac{\pi\qty[2bm + aq(1 + b^2g^2)]}{4\Xi_a \Xi_b^2 G_5} \,, \\
        Q & = \frac{\mathpzc{x} \sqrt{3} \pi q}{4 \Xi_a \Xi_b G_5} \,,
    \end{aligned}
\end{align}
physically corresponding to: energy, angular momenta and electric charge. Above we have defined the constants
\begin{align}\label{eq-casimir-and-zeta}
    E_{\text{AdS}} & = \frac{3\pi}{32 g^2 G_5}\qty(1 + \frac{(\Xi_a - \Xi_b)^2}{9\Xi_a \Xi_b}) \,, \quad \widetilde{\zeta}' = \frac{9g \pi \beta\qty(\Xi_a^2 - \Xi_a \Xi_b + \Xi_b^2)}{\Xi_a \Xi_b G_5}\zeta' \,,
\end{align} 
where $ \zeta'$ is the coefficient of a certain finite counterterm that we introduce to the boundary action in the process of holographic renormalization, see Appendix \ref{sec-on-shell-action}. In a similar calculation \cite{Papadimitriou:2005ii} obtains the energy of the Kerr-$ \text{AdS}_{5}$ black hole --- the special case $ q = 0$, in a renormalization scheme where $ \widetilde{\zeta}' = 0$. For the quantity $ E_{\text{AdS}}$, which physically corresponds to the Casimir energy of empty $ \text{AdS}_{5}$, we find agreement with \cite{Papadimitriou:2005ii}. The Euclidean on-shell action is obtained by evaluating (\ref{eq-action-eucl}), on-shell, on a surface $ r = R_0$, where $ R_0 \gg r_+$. Naively, one obtains divergences, scaling as $ R_0^4$ and $ R_0^2$. They can be cancelled by subtracting the on-shell action of $ \text{AdS}_{5}$, written in suitable coordinates, in a background subtraction procedure. The final answer reads
\begin{align}\label{eq-IrenBS}
    \mathcal{I}_{\text{bs}} & = \frac{\pi\beta}{4\Xi_a \Xi_b G_5} \qty[m - g^2(a^2 + r_+^2)(b^2 + r_+^2) - \frac{q^2 r_+^2}{(a^2 + r_+^2)(b^2 + r_+^2) + abq}] \,.
\end{align}
To obtain this answer, crucially, one has to include the gauge parameter $ \alpha$ in (\ref{eq-cclp-hopf1}) and then set it to $-\Phi$ to ensure that the gauge field is regular on the horizon. This important aspect of the calculation was not explicitly stated in the original derivation presented in \cite{Chen:2005zj}. Alternatively, in Appendix \ref{sec-on-shell-action} we employ holographic renormalization to obtain the following Euclidean on-shell action
\begin{align}\label{eq-IrenHR}
    \mathcal{I} & = \beta \qty(E_{\text{AdS}} - \widetilde{\zeta}') + \mathcal{I}_{\text{bs}} \,,
\end{align}
where precisely the constants (\ref{eq-casimir-and-zeta}) appear. We can interpret this discrepancy between the two on-shell action results as follows. In a non-supersymmetric setting there is no preferred mechanism to fix the coefficient $ \zeta'$ of the finite counterterm, i.e. the value of the constant $ \widetilde{\zeta}'$ defines a choice of renormalization scheme. In particular, the ``background subtraction renormalization scheme'' corresponds to setting $ \widetilde{\zeta}' = E_{\text{AdS}}$. In a different scheme the energy acquires an additive constant factor and the on-shell action acquires the same additive factor times $ \beta$.\\

Having collected the potentials, charges, entropy and on-shell action of CCLP we have verified explicitly that, in any renormalization scheme, the quantum statistical relation holds
\begin{align}\label{eq-qsr}
    \mathcal{I} & = \beta E - S - \beta \Omega_1 J_1 - \beta \Omega_2 J_2 - \beta \Phi Q \,.
\end{align}
Naively, the first law of thermodynamics
\begin{align}\label{eq-first-law}
    \dd{E} & = \frac{1}{\beta} \dd{S} + \Omega_1 \dd{J_1} + \Omega_2 \dd{J_2} + \Phi \dd{Q} \,,
\end{align}
where the variations are taken with respect to the ``bare'' black hole parameters $ (a, \, b, \, r_+, \, q)$, is only satisfied in the renormalization scheme $ \widetilde{\zeta}' = E_{\text{AdS}}$. Insisting on using the ``bare'' parameters, it was argued in \cite{Papadimitriou:2005ii} that in a renormalization scheme where $ \widetilde{\zeta}' \neq E_{\text{AdS}}$ one should use the following generalized first law
\begin{align}
    \dd{E} & = \delta_\sigma E + \frac{1}{\beta} \dd{S} + \Omega_1 \dd{J_1} + \Omega_2 \dd{J_2} + \Phi \dd{Q} \,,
\end{align}
where $ \delta_\sigma E$ is the variation of the energy with respect to a Weyl transformation that keeps representative of the conformal class of boundary metrics fixed. They go further to prove that the variation of the on-shell action under this Weyl transformation is\footnote{Technically, they only prove this in the renormalization scheme $ \widetilde{\zeta}' = 0$, but their result trivially generalizes.}
\begin{align}
    \delta_\sigma \mathcal{I} & = \beta \dd{\qty(E_{\text{AdS}} - \widetilde{\zeta}')} = \beta \, \delta_\sigma E \,.
\end{align}
Using this result, we see that the generalized first law takes the form
\begin{align}\label{eq-gen-first-law}
    \dd{\qty(E - \qty(E_{\text{AdS}} - \widetilde{\zeta}'))} & = \frac{1}{\beta} \dd{S} + \Omega_1 \dd{J_1} + \Omega_2 \dd{J_2} + \Phi \dd{Q} \,,
\end{align}
which is now trivially satisfied in any renormalization scheme, when the variation is taken with respect to the parameters $ (a,b, r_+, q)$.

The Euclidean on-shell action can be though of as minus the logarithm of the grand canonical partition function
\begin{align}
    \mathcal{I}(\beta,\Omega_1,\Omega_2,\Phi) & = - \log \mathcal{Z}(\beta,\Omega_1,\Omega_2,\Phi) \,,
\end{align}
where the grand canonical ensemble is the one in which the thermodynamic potentials $ (\beta, \, \Omega_1, \, \Omega_2, \, \Phi)$ are held fixed, while the charges are allowed to fluctuate. In this setup each state is assigned a probability
\begin{align}
    P_i & = \frac{1}{\mathcal{Z}} e^{-\beta\qty(E_i - \Omega_1 J_{1,i} - \Omega_2 J_{2,i} - \Phi Q_i)} \,, \quad \mathcal{Z} = \sum_i e^{-\beta\qty(E_i - \Omega_1 J_{1,i} - \Omega_2 J_{2,i} - \Phi Q_i)} \,.
\end{align}
Then the average charges are given by
\begin{align}\label{eq-charges-from-thermo}
    \begin{aligned}
        E & = \sum_i P_i E_i = \pdv{\mathcal{I}}{\beta} + \Omega_1 J_1 + \Omega_2 J_2 + \Phi Q \,, \\
        J_1 & = \sum_{i} P_i J_{1,i} = - \frac{1}{\beta} \pdv{\mathcal{I}}{\Omega_1} \,, \quad J_2 = \sum_{i} P_i J_{2,i} = - \frac{1}{\beta} \pdv{\mathcal{I}}{\Omega_2} \,, \\
        Q & = \sum_{i} P_i Q_{i} = - \frac{1}{\beta} \pdv{\mathcal{I}}{\Phi} \,.
    \end{aligned}
\end{align}
The above relations can be readily seen from the quantum statistical relation (\ref{eq-qsr}) and indicate that the potentials $ (\beta, \, \Omega_1, \, \Omega_2, \, \Phi)$ are conjugate to the charges $ (E, \, J_1, \, J_2, \, Q)$, respectively. We can verify these relations on the CCLP solution as follows. First, we express the Euclidean on-shell action and the potentials solely in terms of the black hole parameters $ (a, \, b, \, r_+, \, q)$. Then we solve the following system of four equations
\begin{align}
    \pdv{\mathcal{I}}{(a,b,r_+,q)} & = \pdv{\mathcal{I}}{\beta} \pdv{\beta}{(a,b,r_+,q)} \nonumber\\
    & \qquad + \pdv{\mathcal{I}}{\Omega_1} \pdv{\Omega_1}{(a,b,r_+,q)} + \pdv{\mathcal{I}}{\Omega_2} \pdv{\Omega_2}{(a,b,r_+,q)} + \pdv{\mathcal{I}}{\Phi} \pdv{\Phi}{(a,b,r_+,q)}
\end{align}
for the four unknowns $ \qty(\pdv{\mathcal{I}}{\beta}, \, \pdv{\mathcal{I}}{\Omega_1}, \, \pdv{\mathcal{I}}{\Omega_2}, \, \pdv{\mathcal{I}}{\Phi})$. Plugging the results in (\ref{eq-charges-from-thermo}) we get perfect agreement with the charges obtained in Appendices \ref{sec-charges1} and \ref{sec-charges2} via holographic renormalization and Komar integrals. Finally, one can view the entropy as a function of the charges $ (E, \, J_1, \, J_2, \, Q)$ and write the quantum statistical relation as
\begin{align}
    S(E,J_1,J_2,Q) & = \beta E - \beta \Omega_1 J_1 - \beta \Omega_2 J_2 - \beta \Phi Q - \mathcal{I}(\beta,\Omega_1,\Omega_2,\Phi) \,,
\end{align}
where the following relations should hold
\begin{align}\label{eq-potentials-from-thermo}
    \beta & = \pdv{S}{E} \,, \quad \Omega_1 = - \frac{1}{\beta} \pdv{S}{J_1} \,, \quad \Omega_2 = - \frac{1}{\beta} \pdv{S}{J_2} \,, \quad \Phi = - \frac{1}{\beta} \pdv{S}{Q} \,.
\end{align}
Again, we establish their validity on the CCLP solution by solving the following system of four equations
\begin{align}
    \pdv{S}{(a,b,r_+,q)} & = \pdv{S}{E} \pdv{E}{(a,b,r_+,q)} \nonumber\\
    & \qquad + \pdv{S}{J_1} \pdv{J_1}{(a,b,r_+,q)} + \pdv{S}{J_2} \pdv{J_2}{(a,b,r_+,q)} + \pdv{S}{Q} \pdv{Q}{(a,b,r_+,q)}\,,
\end{align}
for the four unknowns $ \qty(\pdv{S}{E}, \, \pdv{S}{J_1}, \, \pdv{S}{J_2}, \, \pdv{S}{Q})$ and plugging the results in (\ref{eq-potentials-from-thermo}). This analysis shows that the entropy is the Legendre transform of the Euclidean on-shell action, in which the charges $ (E, \, J_1, \, J_2, \, Q)$ replace the potentials $ (\beta, \, \Omega_1, \Omega_2, \Phi)$ as independent variables. 

\subsection{Thermodynamics in the supersymmetric and BPS limits of CCLP}\label{sec-susy-and-bps-limits}

One can make the CCLP solution supersymmetric by saturating the BPS bound
\begin{align}\label{eq-susy-limit}
    q & = \frac{m}{1 + ag + bg} \,.
\end{align}
In what follows it is convenient to also exchange the parameter $ m$ for
\begin{align}\label{eq-mt}
    \widetilde{m} & = \frac{mg}{(a + b)(1 + ag)(1 + bg)(1 + ag + bg)} - 1 \,.
\end{align}
We show that the resulting three-parameter family of solutions, i.e. $ (a, \, b, \, \widetilde{m})$, is supersymmetric in Appendix \ref{sec-killing-spinor} by explicitly solving the Killing spinor equations. This three parameter family of Euclidean supersymmetric solutions can be analytically continued to Lorentzian signature resulting in a family of supersymmetric black holes. These black hole solutions are however not regular and causal outside their horizon which is manifested by the generically complex roots of $ \Delta_r(r) = 0$. Setting $ \widetilde{m} = 0$ one obtains a two-parameter family, i.e. $ (a, \, b)$, of supersymmetric and extremal black holes that do not suffer from such problems and have vanishing temperature, i.e. $ \beta \to \infty$. This was called the BPS limit in \cite{Cabo-Bizet:2018ehj} and one can show that in this BPS limit $ \Delta_r(r) = 0$ has a real double root given by 
\begin{align}
    r_*^2 & = \frac{1}{g}(a + b + abg) \,,
\end{align}
and the surface $ r = r_*$ constitutes the event horizon of the BPS black hole. 

To have better control over the IR divergence associated with the fact that  $ \beta \to \infty$ in the BPS limit, we provide an alternative description of the supersymmetric black hole. Using (\ref{eq-mt}) and (\ref{eq-mTrh}), the supersymmetry condition (\ref{eq-susy-limit}) can be rewritten as
\begin{align}
    q & = -ab + r_+^2(1 + ag + bg) \pm \sqrt{-r_+^2(r_*^2 - r_+^2)} \,.
\end{align}
Insisting that $ q$ is real, we see that we are forced to take the extremal limit $ r_+ \to r_*$ alongside the supersymmetric limit. In general however $ q$ is complex and leads to the following (complex) supersymmetric thermodynamic quantities
\begin{align}\label{eq:potchargesBPS}
        \beta & = - \frac{2\pi(a - \iu r_+)(b - \iu r_+)(gr_*^2 + \iu r_+)}{g(r_*^2 - r_+^2)\qty[2r_+(1 + ag + bg) + \iu g(r_*^2 - 3r_+^2)]}\,, \nonumber\\
        \Omega_1 & = \frac{g(r_*^2 + \iu ar_+)(1 - \iu gr_+)}{(g r_*^2 + \iu r_+)(a - \iu r_+)} \,, \quad \Omega_2 = \eval{\Omega_1}_{a \leftrightarrow b} \,, \nonumber\\
        \Phi & = \frac{\sqrt{3}\, \iu r_+(1 - \iu gr_+)}{\mathpzc{x}(gr_*^2 + \iu r_+)} \,, \nonumber\\
        \widetilde{E} & = \frac{\pi\qty[3 - g^2\qty(a^2(1 + bg) + b^2(1 + ag) - ab)]\qty[-ab + r_+^2\qty(1 + ag + bg) + \iu g r_+\qty(r_*^2 - r_+^2)]}{4\Xi_a \Xi_b(1 - ag)(1 - bg) G_5} \,, \nonumber\\
        J_1 & = \frac{\pi\qty[2a + b(1 + ag)]\qty[-ab + r_+^2\qty(1 + ag + bg) + \iu g r_+\qty(r_*^2 - r_+^2)]}{4\Xi_a \Xi_b(1 - ag)G_5} \,, \quad J_2 = \eval{J_1}_{a \leftrightarrow b} \,, \\
        Q & = \frac{\mathpzc{x}\sqrt{3}\,\pi\qty[-ab + r_+^2\qty(1 + ag + bg) + \iu g r_+\qty(r_*^2 - r_+^2)]}{4\Xi_a \Xi_b G_5} \,, \nonumber\\
        S & = \frac{\pi^2\qty[r_+\qty(r_+^2 + a^2(1 + bg) + b^2(1 + ag) + ab) + \iu abg(r_*^2 - r_+^2)]}{2 \Xi_a \Xi_b G_5}  \nonumber\\
        \widetilde{\mathcal{I}} & = \frac{\pi^2(a - \iu r_+)^2(b - \iu r_+)^2(1 + g^2 r_*^2)}{2\Xi_a \Xi_b\qty[2r_+(1 + ag + bg) + \iu g(r_* - 3r_+^2)]G_5} \,,\nonumber
\end{align}
where we have defined $ \widetilde{E} = E - \qty(E_{\text{AdS}} - \widetilde{\zeta}')$ and $ \widetilde{\mathcal{I}} = \mathcal{I} - \beta\qty(E_{\text{AdS}} - \widetilde{\zeta}')$. As expected, the quantum statistical relation (\ref{eq-qsr}) and the generalized first law (\ref{eq-gen-first-law}) continue to hold in the supersymmetric limit. We also find that the charges obey the following linear relation
\begin{align}\label{eq-susy-algebra}
   E - g J_1 - g J_2 - \frac{\sqrt{3}}{\mathpzc{x}} Q = E_{\text{AdS}} - \widetilde{\zeta}' \,.
\end{align}
If the right hand side of this equation vanishes it will look like the familiar BPS constraint of the charges obeyed by supersymmetric black holes. As explained in Appendix~\ref{sec-CCLP-appendix} the parameter $\widetilde{\zeta}'$ is the coefficient of a finite counterterm in the holographic renormalization procedure used to compute the gravitational charges. We can therefore make the right hand side of \eqref{eq-susy-algebra} vanish if we choose a scheme in which $\widetilde{\zeta}' =E_{\text{AdS}} $. This choice is implemented by the background subtraction scheme discussed above \eqref{eq-IrenBS}. We note that since there is no preferred holographic renormalization scheme for 5d minimal gauged supergravity that is compatible with covariance, gauge invariance, and supersymmetry, see for example \cite{BenettiGenolini:2016tsn}, one is in principle free to choose any other value of the finite counterterm coefficient $\widetilde{\zeta}'$. From now on we will work in the $\widetilde{\zeta}' =E_{\text{AdS}} $ scheme which implies that $ E = \widetilde{E}$ and $ \mathcal{I} = \widetilde{\mathcal{I}}$. 

In addition to the linear relation in (\ref{eq-susy-algebra}), supersymmetry implies a linear relation between the corresponding potentials that reads
\begin{align}\label{eq-susy-constraint1}
    \beta\qty(g + \Omega_1 + \Omega_2 - g \mathpzc{x}\sqrt{3}\, \Phi) & = 2\pi \iu \,.
\end{align}
In Appendix \ref{sec-killing-spinor} we show how this constraint arises from a periodicity condition on the Killing spinor on the horizon. It is useful to define new chemical potentials which are convenient for taking the $\beta \to \infty$ supersymmetric black hole limit
\begin{align}
    \omega_1 & = \beta\qty(\Omega_1 - g) \,, \quad \omega_2 = \beta\qty(\Omega_2 - g) \,, \quad \varphi = \beta\qty(\Phi - \frac{\sqrt{3}}{\mathpzc{x}}) \,.
\end{align}
In terms of them the supersymmetric Euclidean on-shell action takes the remarkably simple form
\begin{align}\label{eq-susy-action-compact}
    \mathcal{I} & = \frac{\pi \mathpzc{x}^3}{12 \sqrt{3}\, G_5} \frac{\varphi^3}{\omega_1 \omega_2} \,.
\end{align}
The quantum statistical relation (\ref{eq-qsr}) and the linear constraint (\ref{eq-susy-constraint1}) can also be written in a manifestly $ \beta$-independent way
\begin{align}\label{eq-susy-qsr}
    \mathcal{I} = - S - \omega_1 J_1 - \omega_2 J_2 - \varphi Q \,; \quad \omega_1 + \omega_2 - g \mathpzc{x} \sqrt{3} \, \varphi & = 2 \pi \iu \,,
\end{align}
where the energy has been eliminated using (\ref{eq-susy-algebra}).

This presentation of the supersymmetric CCLP black hole is adapted to taking the limit to extremality: $ r_+ \to r_*$. Even though $ \beta$ diverges in this limit, the newly defined chemical potentials assume the following finite (and complex) BPS values 
\begin{align}\label{eq:BPSfug}
    \bm{\omega}_1 & = - \frac{\pi(1 - ag)(b - \iu \, r_*)}{r_*\qty[1 + g(a + b - \iu \, r_*)]} \,, \quad \bm{\omega}_2 = \eval{\omega_1}_{a \leftrightarrow b} \,, \quad \bm{\varphi} = \frac{\sqrt{3} \, \pi (a - \iu \, r_*)(b - \iu \, r_*)}{\mathpzc{x} \, r_* \qty[1 + g(a + b - \iu \, r_*)]} \,,
\end{align}
where we adopted the convention to denote BPS quantities in the $\beta \to \infty$ limit with bold letters. The BPS charges and the entropy become real in this limit and read
\begin{align}\label{eq-bps-charges-entropy}
    \begin{aligned}
    \bm{E} & = \frac{\pi(a + b)\qty[3 - g^2\qty(a^2(1 + bg) + b^2(1 + ag) - ab)]}{4g (1 - ag)^2(1 - bg)^2 G_5} \,, \, \, \,  \bm{Q} = \frac{\mathpzc{x}\sqrt{3}\,\pi(a + b)}{4(1 - ag)(1 - bg)G_5} \,, \\
    \bm{J}_1 & = \frac{\pi(a + b)\qty[2a + b(1 + ag)]}{4g(1 - ag)^2(1 - bg)G_5} \,, \quad \bm{J}_2 = \eval{\bm{J}_1}_{a \leftrightarrow b} \,, \quad \bm{S} = \frac{\pi^2(a + b)r_*}{2g(1 - ag)(1 - bg)G_5} \,,\\
    \end{aligned}
\end{align}
while the BPS Euclidean on-shell action remains complex and is naturally expressed in terms of the new BPS potentials
\begin{align}\label{eq-bps-action-compact}
    \bm{\mathcal{I}} & = \mu \frac{\bm{\varphi}^3}{\bm{\omega}_1 \bm{\omega}_2} \,,
\end{align}
where we have extracted the constant pre-factor $ \mu = \pi\mathpzc{x}^3/(12 \sqrt{3} G_5)$. The quantum statistical relation and linear constraint between the potentials take precisely the same form as in the supersymmetric case
\begin{align}\label{eq-bps-qsr}
    \bm{\mathcal{I}} = - \bm{S} - \bm{\omega}_1 \bm{J}_1 - \bm{\omega}_2 \bm{J}_2 - \bm{\varphi} \bm{Q} \,; \quad \bm{\omega}_1 + \bm{\omega}_2 - \nu \bm{\varphi} & = 2 \pi \iu \,,
\end{align}
where we have defined $ \nu = g \mathpzc{x} \sqrt{3}$. We again work in the grand canonical ensemble where the potentials are held fixed and the charges allowed to fluctuate and write the entropy as
\begin{align}\label{eq-entropy-legendre}
    \bm{S}(\bm{J}_1,\bm{J}_2,\bm{Q}) & = - \bm{\mathcal{I}}(\bm{\omega}_1,\bm{\omega}_2,\bm{\varphi}) - \bm{\omega}_1 \bm{J}_1 - \bm{\omega}_2 \bm{J}_2 - \bm{\varphi} \bm{Q} - \lambda\qty(\bm{\omega}_1 + \bm{\omega}_2 - \nu \bm{\varphi} - 2 \pi \iu) \,,
\end{align}
where we have included a Lagrange multiplier $ \lambda$ multiplying the linear constraint. Taking partial derivatives of the above with respect to the potentials one obtains three extremization equations
\begin{align}
    \begin{aligned}
        - \pdv{\bm{\mathcal{I}}}{\bm{\omega}_1} & = \bm{J}_1 + \lambda \,, & - \pdv{\bm{\mathcal{I}}}{\bm{\omega}_2} & = \bm{J}_2 + \lambda \,, & - \pdv{\bm{\mathcal{I}}}{\bm{\varphi}} & = \bm{Q} - \nu \lambda \\
        & & & \Downarrow \\
        \mu \frac{\bm{\varphi}^3}{\bm{\omega}_1^2 \bm{\omega}_2} & = \bm{J}_1 + \lambda \,, & \mu \frac{\bm{\varphi}^3}{\bm{\omega}_1 \bm{\omega}_2^2} & = \bm{J}_2 + \lambda \,, & -\mu \frac{3\bm{\varphi}^2}{\bm{\omega}_1 \bm{\omega}_2^2} & = \bm{Q} - \nu \lambda \,. \\
    \end{aligned}
\end{align}
In principle, to perform the Legendre transform one has to invert the bottom line above (together with the linear constraint) to express $ (\bm{\omega}_1, \bm{\omega}_2, \bm{\varphi}, \lambda)$ in terms of the charges $ (\bm{J}_1, \bm{J}_2, \bm{Q})$ and then plug the results in (\ref{eq-entropy-legendre}) to obtain the entropy. However, due to $ \bm{\mathcal{I}}$ being a homogeneous function of degree $ 1$, from Euler's theorem one has $ \bm{\mathcal{I}} = \bm{\omega}_1 \partial_{\bm{\omega}_1} \bm{\mathcal{I}} + \bm{\omega}_2 \partial_{\bm{\omega}_2} \bm{\mathcal{I}} + \bm{\varphi} \partial_{\bm{\varphi}} \bm{\mathcal{I}}$, and the expression for the entropy simplifies to
\begin{align}
    \bm{S} & = 2\pi \iu \lambda \,.
\end{align}
This provides a shortcut to obtaining the entropy: by inspection we see that
\begin{align}\label{eq-cubic}
    0 = \frac{1}{\nu^3} \qty[ \qty(\pdv{\bm{\mathcal{I}}}{\bm{\varphi}})^3 - 27 \mu \pdv{\bm{\mathcal{I}}}{\bm{\omega}_1} \pdv{\bm{\mathcal{I}}}{\bm{\omega}_2} ] \implies 0 = p_0 + p_1 \lambda + p_2 \lambda^2 + \lambda^3 \,,
\end{align}
where after the implies sign we have used the extremization equations and extracted the coefficients in front of the $ \lambda$ terms
\begin{align}
    p_0 & = - \frac{1}{\nu^3} \bm{Q}^3 - \frac{27\mu}{\nu^3} \bm{J}_1 \bm{J}_2 \,,  \quad p_1 = \frac{3}{\nu^2} \bm{Q}^2 - \frac{27 \mu}{\nu^3}(\bm{J}_1 + \bm{J}_2) \,, \quad p_2 = - \frac{3}{\nu} \bm{Q} - \frac{27\mu}{\nu^3} \,.
\end{align}
We seek a purely real entropy in the BPS limit, so we should have a purely imaginary root of the cubic (\ref{eq-cubic}). This is possible if $ p_0 = p_1 p_2$, which translates to the following non-linear relation between the BPS charges:
\begin{align}\label{eq:cubicconstr}
    \qty(\frac{2}{g \mathpzc{x} \sqrt{3}}\bm{Q})^3 + \frac{2\pi}{g^3 G_5} \bm{J}_1 \bm{J}_2 & = \qty(\frac{2 \sqrt{3}}{g \mathpzc{x}} \bm{Q} + \frac{\pi}{2 g^3 G_5 })\qty[\qty(\frac{2}{g \mathpzc{x}} \bm{Q})^2 - \frac{\pi}{g^3 G_5}\qty(\bm{J}_1 + \bm{J}_2)] \,.
\end{align}
Then the purely imaginary roots are $ \lambda = \pm \iu \, \sqrt{p_1}$. Picking the minus sign in order to have positive entropy we obtain
\begin{align}\label{eq:SCCLPsusy}
    \bm{S} & = \frac{\pi}{g} \sqrt{ \frac{4}{\mathpzc{x}^2} \bm{Q}^2 - \frac{\pi}{g G_5}\qty(\bm{J}_1 + \bm{J}_2)} \,.
\end{align}
While we have shown how to obtain the supersymmetric black hole entropy as a Legendre transformation of the on-shell action it is also possible to verify the validity of \eqref{eq:SCCLPsusy} and \eqref{eq:cubicconstr} using the explicit expression of the BPS charges and entropy in terms of the black hole parameters given in (\ref{eq-bps-charges-entropy}).

\section{Holography and the superconformal index}\label{sec-holography}

The results for the on-shell action and entropy of the supersymmetric CCLP solution presented above can be understood from a holographic viewpoint in terms of an appropriate supersymmetric partition function on $S^1\times S^3$ called the superconformal index $\mathcal{I}_{\rm QFT}$, see \cite{Rastelli:2016tbz} for a review. Here we discuss how this can be done in some detail.

\subsection{Anomalies and the index on the second sheet}\label{subsec-2ndsheet}

The superconformal index of class $\mathcal{S}$ SCFTs has been studied for low values of $N$ in \cite{Beem:2012yn,Gadde:2009kb,Gadde:2011ik,Rastelli:2014jja}. We are not aware of an extension of these results valid in the large $N$ limit relevant to our holographic setup. Recently, building on previous results in \cite{Hosseini:2017mds,Cabo-Bizet:2018ehj}, it was shown in \cite{Cassani:2021fyv},  see also \cite{GonzalezLezcano:2020yeb}, that the superconformal index of general 4d $\mathcal{N}=1$ SCFTs is given by the following compact expression
\begin{equation}\label{eq:CK2ndsheet}
\begin{split}
\log \mathcal{I}_{\rm QFT} &= \frac{\varphi_R^3}{48\omega_1\omega_2} k_{RRR} -\frac{\varphi_R(\omega_1^2+\omega_2^2-4\pi^2)}{48\omega_1\omega_2}  k_R +\log |G| \,,
\end{split}
\end{equation}
where, $\omega_{1,2}$ are the fugacities for the two angular momenta on $S^3$, $\varphi_R$ is the $U(1)$ R-symmetry fugacity, $G$ is the order of the Abelian 1-form symmetry in the theory (if any), and $(k_{RRR},k_R)$ are the cubic and linear 't Hooft anomalies. Supersymmetry imposes the following linear relation between the fugacities 
\begin{equation}\label{eq:varphiRrel}
\varphi_R = (\omega_1+\omega_2+2\pi {\rm i} n_0)\,,
\end{equation}
and the expression in \eqref{eq:CK2ndsheet} is valid for $n_0=\pm 1$, i.e. on the ``second sheet'' of the complexified fugacities. This result for the index includes the contribution from the supersymmetric Casimir energy \cite{Bobev:2015kza,Assel:2015nca} and is valid up to exponentially suppressed terms of the form $e^{-\ell_3/\beta}$ where $\ell^3$ is the radius of the $S^3$ and $\beta$ is the circumference of the $S^1$. It is not immediately clear that this Cardy-like limit of the index is useful for studying its large $N$ properties. However, it was shown in \cite{Hosseini:2017mds,Cabo-Bizet:2018ehj,Bobev:2022bjm,Cassani:2022lrk} that the $k_{RRR}$ and $k_R$ terms in \eqref{eq:CK2ndsheet} are in perfect agreement with the regularized on-shell action of the supersymmetric CCLP solution in 5d $\mathcal{N}=2$ minimal gauged supergravity and its four-derivative extension. This in turn implies that the on-shell action and entropy of the wrapped M5-brane black holes we constructed above can be reproduced by using the index in \eqref{eq:CK2ndsheet}. This is indeed possible as we show below.

To use the expression in \eqref{eq:CK2ndsheet} we need the 't Hooft anomalies for the R-symmetry of the SCFT at hand. The 4d $\mathcal{N}=1$ class $\mathcal{S}$ SCFTs of interest here have a $U(1)_F$ flavor symmetry in addition to the $U(1)_R$ R-symmetry. The 't Hooft anomalies for these global symmetries can be computed as in \cite{Gaiotto:2009gz,Benini:2009mz,Alday:2009qq,Bah:2011vv,Bah:2012dg} either with field theory methods or by integrating the anomaly polynomial of the 6d $\mathcal{N}=(2,0)$ SCFT over $\Sigma_{\mathfrak{g}}$. For concreteness here we focus on the type $A_{N-1}$ class $\mathcal{S}$ theories and take $\mathfrak{g}>1$\footnote{The results below have a straightforward generalization to arbitrary genus and to the type $D_{N}$ theories.} for which the 't Hooft anomalies read
\begin{align}\label{eq:kRRRBBBW}
\begin{aligned}
k_{RRR} &= \frac{2(\mathfrak{g}-1)}{27z^2}\left[9z^2-1+(3z^2+1)^{\frac{3}{2}}\right]N^3\,, & k_{RRF} & = \frac{(\mathfrak{g}-1)}{9}\, z\, N\,, \\
k_{RFF} &= -\frac{(\mathfrak{g}-1)}{3}\sqrt{3z^2+1} N^3 \,, & k_{FFF} & = (\mathfrak{g}-1)\, z\, N^3\,, \\
k_{R}  & = \frac{\mathfrak{g}-1}{3}\left[4-\sqrt{3z^2+1}\right]N\,, & k_F & = (\mathfrak{g}-1)\, z\, N \,.
\end{aligned}
\end{align}
We have presented the results for the leading term in the large $N$ expansion of the anomalies which is all we need for our holographic discussion. The exact finite $N$ values can be easily found using the results in \cite{Bah:2011vv,Bah:2012dg}. To obtain these results for the 't Hooft anomalies, as explained in \cite{Bah:2011vv,Bah:2012dg}, one has to employ $a$-maximization \cite{Intriligator:2003jj}. As expected from $a$-maximization we find that $k_{RFF}$ is negative and that the superconformal Ward identity $k_F = 9 k_{RRF}$ is obeyed. In addition, one can use superconformal Ward identities to find the following conformal anomalies
\begin{equation}
\begin{split}
a &= \frac{1}{32}(9 k_{RRR}-3k_R)= (\mathfrak{g}-1)\frac{\left(9z^2-1+(1+3z^2)^{\frac{3}{2}}\right)}{48z^2} N^3\,,\\
c&= \frac{1}{32}(9 k_{RRR}-5k_R)=(\mathfrak{g}-1)\frac{\left(9z^2-1+(1+3z^2)^{\frac{3}{2}}\right)}{48z^2} N^3\,.
\end{split}
\end{equation}
Again we have presented the conformal anomalies only to leading order in the large $N$ limit and as expected for a holographic SCFT we find that $a=c$ to this order. 

Plugging the results for $k_{RRR}$ and $k_R$ presented above in the formula for the superconformal index \eqref{eq:CK2ndsheet} we obtain an explicit expression for $\mathcal{I}_{\rm QFT}$ valid in the Cardy-like limit and to leading order in the large $N$ limit. To compare $\mathcal{I}_{\rm QFT}$ with the regularized on-shell action of the supersymmetric CCLP solution we need to employ the holographic dictionary. As shown in \cite{Bobev:2021qxx,Bobev:2022bjm} the cubic 't Hooft anomaly is related to the 5d AdS radius and Newton constant as
\begin{equation}
k_{RRR} = \frac{4\pi L^3}{9 G_5}\,,
\end{equation}
while $k_R$ is related to the coefficients of the four-derivative corrections to the 5d supergravity theory. To relate this 5d supergravity expression for $k_{RRR}$ to the microscopic parameters $(N,\mathfrak{g},z)$ one should use the relation between the 5d and 7d Newton constants in \eqref{eq-G7-G5-relation}, the expressions that define the AdS$_5$ supersymmetric vacua in \eqref{eq-constants-ansatz}, as well as the uplift of this solution to 11d using the formulae in Section~\ref{sec-M-to-7}. As shown in \cite{Bobev:2021qxx,Bobev:2022bjm}, and can be checked explicitly using the formulae above, this supergravity calculation leads to exactly the same result for $k_{RRR}$ as the one in \eqref{eq:kRRRBBBW}. 

Another entry from the holographic dictionary we need is the relation between the fugacities specifying the index and the gravitational potentials of the CCLP solution. This was also discussed in some detail recently in \cite{Bobev:2022bjm} where it was shown that the supergravity BPS potentials in \eqref{eq:BPSfug} can be related to the field theory fugacities as follows
\begin{equation}
\omega_{1,2} \longleftrightarrow \bm{\omega}_{1,2}\,, \qquad\qquad \varphi_R \longleftrightarrow \frac{\mathpzc{x} \sqrt{3}}{L} \bm{\varphi}\,.
\end{equation}
Comparing the supergravity BPS relation in \eqref{eq-bps-qsr} with the field theory one in \eqref{eq:varphiRrel} we see that we must set $n_0=-1$ in order to find agreement. This choice for the sign of $n_0$ is related to a conventional choice in the sign of the electric field in the CCLP solution.

Combining all these ingredients together we find that regularized supergravity on-shell action $\bm{\mathcal{I}}$ in \eqref{eq-bps-action-compact} precisely agrees with the $k_{RRR}$ term in the logarithm of the superconformal index $\log \mathcal{I}_{\rm QFT}$ in \eqref{eq:CK2ndsheet}. If one would like to obtain the entropy of the supersymmetric CCLP black hole in \eqref{eq:SCCLPsusy} one can simply take $\log \mathcal{I}_{\rm QFT}$ and repeat the Legendre transformation procedure discussed above \eqref{eq:SCCLPsusy}. These results constitute a precision test of holography for the wrapped M5-brane black holes we constructed above and a microscopic understanding of their Bekenstein-Hawking entropy in terms of the superconformal index of the dual 4d $\mathcal{N}=1$ class $\mathcal{S}$ SCFTs.

\subsection{More general black holes?}\label{subsec-generalBH}

The 4d $\mathcal{N}=1$ SCFTs in \cite{Bah:2011vv,Bah:2012dg} have a $U(1)_F$ flavor symmetry in addition to the $U(1)_R$ R-symmetry. This means that the superconformal index can be additionally refined with a flavor fugacity $\varphi_F$. The explicit form of this refined index in the Cardy-like or large $N$ limit has not been studied in the literature. From the supergravity point of view the $U(1)_F\times U(1)_R$ global symmetry is realized by the two Killing vectors in the 11d metric in \eqref{eq:S4metric} or alternatively by the two Abelian gauge fields in the 7d gauged supergravity \eqref{7DSUGRAaction}. The asymptotically AdS$_5$ black hole solutions we studied above have only one electric charge associated with the $U(1)_R$ gauge field. Both the QFT and supergravity setups therefore suggest that there may be more general black hole solutions with two electric charges and two angular momenta with entropy and on-shell action that can be accounted for by the large $N$ limit of the refined superconformal index. While we have not calculated the refined index or constructed these supergravity solutions we present some speculations and educated guesses about their properties. 

We begin with the following proposal for the generalized version of the ``second sheet'' formula for the index of a general 4d $\mathcal{N}=1$ SCFT with a $U(1)$ flavor symmetry
\begin{equation}\label{eq:CKgen2ndsheet}
\begin{split}
\log \mathcal{I}_{\rm QFT} &= \frac{k_{RRR}\,\varphi_R^3+k_{RRF}\,\varphi_R^2\varphi_F+k_{RFF}\,\varphi_R\varphi_F^2+k_{FFF}\,\varphi_F^3}{48\omega_1\omega_2} \,,
\end{split}
\end{equation}
where $\varphi_R$ obeys the linear constraint \eqref{eq:varphiRrel} and we have only presented terms that involve the cubic 't Hooft anomalies since these should be the leading contributions to the large $N$ limit of the index of a holographic SCFT. This proposed formula for the index is based on the form of the supersymmetric Casimir energy for general 4d $\mathcal{N}=1$ SCFTs in \cite{Bobev:2015kza} as well as the holographic results for the index of $\mathcal{N}=4$ SYM in \cite{Hosseini:2017mds}.

Then, we propose the following identification between the field theory fugacities and the supergravity BPS potentials
\begin{align}
    \omega_{1,2} & \longleftrightarrow \bm{\omega}_{1,2} \,, \quad \varphi_R \longleftrightarrow \nu \bm{\varphi}_R \,, \quad \varphi_F \longleftrightarrow \rho \bm{\varphi}_F \,,
\end{align}
where, the constants $ \nu$ and $ \rho$ should be determined by using the holographic dictionary using the explicit putative black hole solutions.To be consistent with the form of the on-shell action in \eqref{eq-bps-action-compact} we expect that $ \nu = \mathpzc{x} \sqrt{3} \, g$, such that $ \bm{\varphi} \equiv \bm{\varphi}_R$.

In the dual supergravity there should be a supersymmetric Euclidean solution with two angular momenta and two electric charges that has the following on-shell action
\begin{align}\label{eq:Ionshellguess}
    \bm{\mathcal{I}} & = \frac{\mu_1 \, \bm{\varphi}_R^3 + \mu_2 \, \bm{\varphi}_R^2 \bm{\varphi}_F + \mu_3 \, \bm{\varphi}_R \bm{\varphi}_F^2 + \mu_4 \, \bm{\varphi}_F^3}{\bm{\omega}_1 \bm{\omega}_2} \,,
\end{align}
where the real parameters $ \mu_{1,2,3,4}$ should be determined by the coupling constants of the particular supergravity theory hosting the proposed new black hole. One should then derive, via the holographic dictionary, explicitly how they are related to the 't Hooft anomalies $ k_{R R R}, k_{R R F}, k_{R F F}$ and $ k_{F F F}$. In order to have a holographic match between $ \bm{\mathcal{I}}$ and $ \log \mathcal{I}_{\text{QFT}}$ we need the following relations to hold
\begin{align}\label{eq-mu-in-terms-of-k}
    \mu_1 & = \frac{k_{RRR}\nu^3}{48} \,, \quad \mu_2 = \frac{k_{RRF} \nu^2 \rho}{48} \,, \quad \mu_3 = \frac{k_{RFF} \nu \rho^2}{48} \,, \quad \mu_4 = \frac{k_{FFF}\rho^3}{48} \,,
\end{align}
Establishing this equivalence is of course non-trivial and hinges on the explicit construction of the black hole solution with two electric charges, the details of the holographic map, and the validity of the formula for the index in \eqref{eq:CKgen2ndsheet}.

If all our educated guesses \eqref{eq:CKgen2ndsheet}-\eqref{eq-mu-in-terms-of-k} are correct and such a non-trivial agreement can indeed be established one can try to proceed and perform a Legendre transform of the on-shell action, or equivalently the large $N$ index in order to derive the entropy of the black hole. This proves to be a non-trivial calculation as we now outline. First let us rewrite the on-shell action as \begin{align}\label{eq-action-tensor}
    \bm{\mathcal{I}} & = \frac{\pi}{24 \bm{\omega}_1 \bm{\omega}_2} \kappa_{IJK} \bm{\varphi}^I \bm{\varphi}^J \bm{\varphi}^K \,,
\end{align}
where $ \kappa_{IJK}$ is totally symmetric with its independent components given by
\begin{align}
    \kappa_{111} & = \frac{24\mu_1}{\pi} \,, \quad \kappa_{112} = \frac{8\mu_2}{\pi} \,, \quad \kappa_{122} = \frac{8\mu_3}{\pi} \,, \quad \kappa_{222} = \frac{24\mu_4}{\pi} \,,
\end{align}
and $ \bm{\varphi}^I = (\bm{\varphi}_R, \bm{\varphi}_F)$. This rewriting of the on-shell action is inspired by the on-shell action of supersymmetric black holes in 5d gauged supergravity coupled to two vector multiplets studied in \cite{Cassani:2019mms}. The only difference here is that we use a basis of supergravity BPS potentials that explicitly singles out the superconformal $ R$-symmetry. To find the entropy we now follow the familiar procedure of extremizing
\begin{align}
    \bm{S} & = - \bm{\mathcal{I}} - \bm{\varphi}^I \bm{Q}_I - \bm{\omega}_1 \bm{J}_1 - \bm{\omega}_2 \bm{J}_2 - \lambda (\bm{\omega}_1 + \bm{\omega}_2 - \nu \bm{\varphi}_R - 2\pi \iu) \,,
\end{align}
where $ \bm{Q}_I = (\bm{Q}_R,\bm{Q}_F)$ and, crucially, only $ \bm{\varphi}_R$ is on the ``second sheet''. The extremization equations read
\begin{align}\label{eq-new-expremization-equs}
    -\pdv{\bm{\mathcal{I}}}{\bm{\omega}_1} & = \bm{J}_1 + \lambda \,, \quad -\pdv{\bm{\mathcal{I}}}{\bm{\omega}_2} = \bm{J}_2 + \lambda \,, \quad - \pdv{\bm{\mathcal{I}}}{\bm{\varphi}_R} = \bm{Q}_R - \nu \lambda \,, \quad - \pdv{\bm{\mathcal{I}}}{\bm{\varphi}_F} = \bm{Q}_F \,.
\end{align}
Together with the linear constraint $ \bm{\omega}_1 + \bm{\omega}_2 - \nu \bm{\varphi}_R = 2\pi \iu$, this amounts to a system of  $ 5$ non-linear equations that need to be solved for  $ (\bm{\omega}_1,\bm{\omega}_2,\bm{\varphi}_R,\bm{\varphi}_F,\lambda)$ in terms of the charges $ (\bm{J}_1,\bm{J}_2,\bm{Q}_R,\bm{Q}_F)$. We have not been able to solve this system of equations analytically and have only found numerical solutions. To proceed further we resort to the shortcut outlined in Section \ref{sec-susy-and-bps-limits}. We note that $ \bm{\mathcal{I}}$ is again homogeneous of degree $ 1$ and the extremized entropy is therefore given by $ \bm{S} = 2\pi \iu \lambda$. 

We note that the relation
\begin{align}\label{eq-guessed-zero}
    0 & = \frac{1}{6} \kappa^{IJK} \pdv{\bm{\mathcal{I}}}{\bm{\varphi}^I} \pdv{\bm{\mathcal{I}}}{\bm{\varphi}^J} \pdv{\bm{\mathcal{I}}}{\bm{\varphi}^K} - \frac{\pi}{4} \pdv{\bm{\mathcal{I}}}{\bm{\omega}_1} \pdv{\bm{\mathcal{I}}}{\bm{\omega}_2} \,,
\end{align}
holds provided that $ \kappa_{IJK}$ obeys
\begin{align}\label{eq-sugra-relation}
    \kappa_{IJK}\kappa_{J'(LM}\kappa_{PQ)K'} g^{JJ'} g^{KK'} & = \frac{4}{3} g_{I(L}\kappa_{MPQ)} \,,
\end{align}
where the metric and the inverse metric read
\begin{align}
    g_{IJ} & = 6 \qty(\frac{\pi}{2 \mu_1})^{-2/3} \delta_{IJ} \,, \quad g^{IJ} = \frac{1}{6} \qty(\frac{\pi}{2 \mu_1})^{2/3} \delta^{IJ} \,,
\end{align}
and consequently, the non-zero components of $ \kappa^{IJK}$ are given by
\begin{align}
    \kappa^{111} & = \frac{\pi}{18 \mu_1} \,, \quad \kappa^{112} = \frac{\pi\mu_2}{54 \mu_1^2} \,, \quad \kappa^{122} = \frac{\pi\mu_3}{54 \mu_1^2} \,, \quad \kappa^{222} = \frac{\pi\mu_4}{18 \mu_1^2} \,.
\end{align}
In the symmetric fugacity basis used in the supergravity analysis in \cite{Cassani:2019mms} the relation (\ref{eq-sugra-relation}) holds since it ensures that the scalars residing in the two vector multiplets parametrize a symmetric coset space. If we take this requirement imposed by gauged supergravity as a given and use the extremization equations (\ref{eq-new-expremization-equs}) together with (\ref{eq-guessed-zero}) we find that $ \lambda$ has to obey the following cubic equation:
\begin{align}
    \begin{aligned}
        0 & = p_0 + p_1 \lambda + p_2 \lambda^2 + \lambda^3 \,, \\
        p_0 & = - \frac{1}{\nu^3} \bm{Q}_R^3 - \frac{27\mu_1}{\nu^3} \bm{J}_1 \bm{J}_2 - \frac{\mu_2}{\mu_1 \nu^3} \bm{Q}_R^2 \bm{Q}_F - \frac{\mu_4}{\mu_1 \nu^3} \bm{Q}_F^3 \,, \\
        p_1 & = - \frac{27\mu_1}{\nu^3}(\bm{J}_1 + \bm{J}_2) + \frac{3}{\nu^2} \bm{Q}_R^2 + \frac{2\mu_2}{\mu_1 \nu^2} \bm{Q}_R \bm{Q}_F + \frac{\mu_3}{\mu_1 \nu^2} \bm{Q}_F^2 \,, \\
        p_2 & = - \frac{27\mu_1}{\nu^3} - \frac{3}{\nu} \bm{Q}_R - \frac{\mu_2}{\mu_1 \nu} \bm{Q}_F \,.
    \end{aligned}
\end{align}
Then the non-linear relation between the charges and the entropy read
\begin{align}
    p_0 & = p_1 p_2 \,, \quad \bm{S} = 2\pi \sqrt{p_1} \,.
\end{align}
Note that these expression nicely reduce to our results in Section \ref{sec-susy-and-bps-limits} when $ \bm{Q}_F = 0$ and 
\begin{align}
    \mu_1 & = \frac{k_{RRR}\nu^3}{48} = \frac{1}{48} \frac{4\pi L^3}{9G_5} \qty(\frac{\mathpzc{x}\sqrt{3}}{L})^3 = \frac{\pi\mathpzc{x}^3}{12 \sqrt{3} G_5} = \mu \,,
\end{align}
where we have used our educated guess for the relation between the supergravity constant $ \mu_1$ and the SCFT anomaly $ k_{R R R}$. 
More importantly, this expression for the entropy and the non-linear constraint between the black hole charges are only valid when (\ref{eq-sugra-relation}) is satisfied. From the point of view of the dual SCFT the origin of this relation is far from clear. We assume that $ \mu_2$ is subleading in large $ N$, which is indeed realized by the concrete example of the class $\mathcal{S}$ theories studied above, see \eqref{eq-mu-in-terms-of-k} and \eqref{eq:kRRRBBBW} where $ \rho$ is assumed not to scale with $ N$. Using this we find that the only way to satisfy the constraint in (\ref{eq-sugra-relation}) is if the following relations hold
\begin{align}\label{eq:constrmu}
    \mu_3 & = - \frac{3}{2}\mu_1 \,, \quad \mu_4 = \frac{1}{\sqrt{2}} \mu_1 \,.
\end{align}
We are not aware of any QFT reason why such a relation should be obeyed by the supergravity constants $ \mu_1, \mu_3$ and $ \mu_4$. Indeed, using our educated guess \eqref{eq-mu-in-terms-of-k}, in terms of the QFT anomalies these relations read
\begin{align}
    k_{RFF}\rho^2 & = - \frac{3}{2} k_{RRR}\nu^2 \,, \quad k_{FFF}\rho^3 = \frac{1}{\sqrt{2}} k_{RRR}\nu^3
\end{align}
Clearly, these relations cannot hold for the explicit anomalies \eqref{eq:kRRRBBBW} for any value of $ \rho$ , even if we allow it to be $ z$-dependent.

Based on the discussion above we conclude that for general holographic SCFTs with a $U(1)_F$ flavor symmetry, including the particular example of the $\mathcal{S}$ SCFTs discussed above, one cannot find a simple compact expression for the black hole entropy in terms of the two electric charges and two angular momenta. This points to the fact that perhaps the sought more general wrapped M5-brane black holes with an additional flavor electric charge cannot be constructed using a 5d $\mathcal{N}=2$ gauged supergravity theory coupled to a finite number of vector multiplets. Indeed, based on the consistent truncation classification results in \cite{Cassani:2020cod,Josse:2021put} there are no candidates for such a gauged supergravity theory.

The situation is different for $\mathcal{N}=4$ SYM. One can show that for this theory the 't Hooft anomalies for the $U(1)^3$ Cartan subalgebra of the $SO(6)$ global symmetry obey the constraints in \eqref{eq:constrmu} and indeed, as shown in \cite{Kim:2006he}, there is a simple compact expression for the entropy of the supersymmetric black hole with two angular momenta and three electric charges constructed in \cite{Kunduri:2006ek} using the STU model of 5d $\mathcal{N}=2$ gauged supergravity.

\section{Discussion}\label{sec-Discussion}

The results presented above leave several open questions and interesting possibilities for generalizations. Here we discuss some of them.

\begin{itemize}

\item As discussed in Section~\ref{subsec-generalBH} it is very likely that there are more general supersymmetric black holes with an additional electric charge corresponding to the $U(1)_F$ flavor symmetry of the class $\mathcal{S}$ SCFTs. It will be very interesting to construct these solutions, either by using an appropriate 5d matter-coupled supergravity theory arising as a consistent truncation of 11d supergravity, or by a judicious Ansatz for the fields of the 7d maximal gauged supergravity. Of course it should in principle be possible to construct such solutions directly in 11d supergravity but we expect that to be prohibitively difficult since the corresponding BPS equations should reduce to a system of coupled nonlinear PDEs.

\item There are two other generalizations of the black hole solutions discussed above that one can contemplate. For concreteness here we have focused the discussion on the class $\mathcal{S}$ SCFTs associated to the 6d $\mathcal{N}=(2,0)$ SCFT of type $A_{N-1}$ compactified on a smooth $\Sigma_{\mathfrak{g}}$ with $\mathfrak{g}>1$. Using the results in \cite{Bah:2011vv,Bah:2012dg} it is straightforward to generalize this setup to the $D_N$ type class $\mathcal{S}$ SCFTs and to $\mathfrak{g}=0,1$. On the supergravity side this corresponds to modifying the 11d supergravity solutions discussed above to have an internal $\mathbb{RP}^4$ space instead of the $S^4$ used in \eqref{eq:S4metric} and to modify the metric in \eqref{eq:H2met} to that of the torus or the two-sphere. More non-trivially, one can attempt to find a larger class of black hole solutions arising from wrapped branes on punctured Riemann surfaces. While it is known how to construct such AdS$_5$ vacua with $\mathcal{N}=2$ and $\mathcal{N}=1$ supersymmetry, see \cite{Gaiotto:2009gz,Bah:2013qya,Bah:2015fwa}, it is not entirely clear how to generalize these supergravity solutions to include a black hole. Perhaps the 7d gauged supergravity method developed in \cite{Bobev:2019ore} to treat some special punctures on the Riemann surface will prove useful in this regard. 

\item As already emphasized in Section~\ref{subsec-2ndsheet} it is not entirely clear to us why the Cardy-like limit of the index used to derive the ``second sheet'' formula for the index in \cite{Cassani:2021fyv} has a larger regime of validity that captures the behavior of the large $N$ limit of the index relevant for holography. It will be most interesting to understand this better and also to generalize the ``second sheet'' formula for the index to include fugacities associated with flavor symmetries. We have proposed such a generalization in \eqref{eq:CKgen2ndsheet} and it will be interesting to check our educated guess more rigorously. Finally, it is important to study $\log N$ corrections to the large $N$ class $\mathcal{S}$ superconformal index which should correspond to the 1-loop contributions to the gravitational path integral of the KK modes of 11d supergravity around the AdS$_5$ black hole solutions. Establishing this relation rigorously will provide a stringent precision test of holography. We also note in passing that subleading terms in the large $N$ power law expansion of the superconformal index are also captured by ``second sheet'' formula. These correspond to higher-derivative corrections to the CCLP solution in gauged supergravity. This was studied recently in \cite{Bobev:2022bjm,Cassani:2022lrk} where a precise agreement between the holographic and QFT results was established. In particular a discussion on higher-derivative corrections to AdS$_5$ black holes arising from wrapped branes was presented in \cite{Bobev:2022bjm} based on the assumptions that such solutions indeed exist. Our results here indeed show that this is true.

\item The superconformal index for the class $\mathcal{S}$ SCFTs discussed above was studied in \cite{Beem:2012yn} for low values of $N$. It will be most interesting to understand how to extend the results of this work to larger values of $N$. For gauge theories coupled to matter a useful approach to studying the large $N$ limit of the index is to employ the Bethe Ansatz re-formulation, see \cite{Benini:2018mlo,Benini:2018ywd}. If such a formulation is possible for class $\mathcal{S}$ SCFTs it will provide a concrete calculational tool to gain valuable insights into the quantum gravity corrections to the properties of the black holes constructed in this work. 

\end{itemize}

\bigskip
\section*{Acknowledgments}
We are grateful to Chris Beem, Davide Cassani, Fri\dh rik Freyr Gautason, Junho Hong, Kiril Hristov, Emanuel Malek, Dario Martelli,  and especially Valentin Reys for valuable discussions.  We are supported in part by an Odysseus grant G0F9516N from the FWO and by the KU Leuven C1 grant ZKD1118 C16/16/005. VD and AV are also supported by doctoral fellowships with numbers 11C8422N (VD) and 1102722N (AV) from the Research Foundation - Flanders (FWO).


\newpage
\appendix

\section{Details on the dimensional reduction from 7d to 5d}\label{appendix7d5d}
\subsection{Reducing the action}\label{proofappendix}
We begin with the 7d action (\ref{7DSUGRAaction}) and set the parameter $ \mu = 2/L$, where $ L$ is eventually going to be the $ \text{AdS}_{5}$ radius. We define the respective volume forms
\begin{align}
    \varepsilon_{(7)} & =  \star_7 1 \,, \quad \varepsilon_{(5)} =  \star_5 1 \,, \quad \varepsilon_{(2)} = \star_2 1 = \dd{\omega} \,,
\end{align}
and use the Ansatz in \eqref{eq-ansatz} to calculate
\begin{align}
    \varepsilon_{(7)} & = e^{5f_0 + 2g_0}L^2 \, \varepsilon_{(5)} \wedge \dd{\omega} \,.
\end{align}
Then we express the constituents of the 7d action purely in terms of the 5d fields and 5d Hodge star
\begin{align}
    \begin{aligned}
        S & = \mathpzc{c} \, e^{2f_0 + 2\Lambda_+} \star_5 F \,, \\
        \star_7 S & = -\mathpzc{c} \, e^{f_0 + 2g_0 + 2\Lambda_+}L^2 \, F \wedge \dd{\omega} \,, \\
        F_{1,2} & = \mathpzc{a} \, (1 \pm z)L \, \dd{\omega} + \mathpzc{b} \, e^{f_0 + 2\Lambda_{1,2}} \, F \,, \\
        \star_7 F_{1,2} & = \mathpzc{a} \, e^{5f_0 - 2g_0}(1 \pm z) L^{-1} \, \varepsilon_{(5)} + \mathpzc{b} \, e^{2f_0 + 2g_0 + 2\Lambda_{1,2}} L^2 \, \star_5 F \wedge \dd{\omega} \,.
    \end{aligned}
\end{align}
Finally, we evaluate the 7d Ricci scalar in terms of the 5d Ricci scalar
\begin{align}
    R_{(7)} & = e^{-2f_0} R_{(5)} + e^{-2g_0} L^{-2} R_{(2)} \,, \quad R_{(2)} = -2 \,.
\end{align}
These expressions allow us to write down all the terms in the 7d action as
\begin{align}\nonumber
    \begin{aligned}
        \star_7 R & = -2 \, e^{5f_0} \, \varepsilon_{(5)} \wedge \dd{\omega} + L^2_0 \, \star_5 R \wedge \dd{\omega} \,, \\
        -  \star_7 V & = - e^{5f_0 + 2g_0} L^2 \, V \, \varepsilon_{(5)}\wedge \dd{\omega} \,, \\
        - \frac{1}{2}e^{-4\Lambda_{1,2}} F_{1,2} \wedge \star_7 F_{1,2} & = - \frac{\mathpzc{a}^2}{2} \, e^{5f_0-2g_0 - 4\Lambda_{1,2}} (1 \pm z)^2 \, \varepsilon_{(5)} \wedge \dd{\omega} - \frac{\mathpzc{b}^2}{2} \, L^2_0 \, F \wedge  \star_5 F \wedge \dd{\omega} \,, \\
        - \frac{1}{2} e^{-4\Lambda_+} S \wedge  \star_7 S & = \frac{\mathpzc{c}^2}{2} \, L^2_0 \, F \wedge \star_5 F \wedge \dd{\omega} \,, \\
        \frac{L}{4} S \wedge \dd{S} & = 0 \,, \\
        - \frac{L}{2} F_1 \wedge F_2 \wedge S & = -4 \mathpzc{a}\mathpzc{b}\mathpzc{c} \, L_0^2 \, F \wedge \star_5 F \wedge \dd{\omega} \,, \\
        \frac{L}{8}\qty[A_1 \wedge F_1 \wedge F_2 \wedge F_2 + (1 \leftrightarrow 2)] & = 4 \mathpzc{a} \mathpzc{b}^3 \, L_0^2 \, A \wedge F \wedge F \wedge \dd{\omega}\,.
     \end{aligned}
\end{align}
Here we have defined $ L_0^2 \equiv e^{3f_0 + 2g_0}L^2$ and in the last expression we integrated by parts. The terms on the right hand side proportional to $ L_0^2 \, \varepsilon_{(5)} \wedge \dd{\omega}$ sum to 
\begin{align}
    - \frac{1}{2L^2} e^{2f_0 - 2g_0 - 2\Lambda_+}\qty(2 e^{2g_0 + 4\Lambda_+}\qty(2 + e^{2g_0}L^2 V) + \mathpzc{a}^2\qty(e^{4\Lambda_1}(1 - z)^2 + e^{4\Lambda_2}(1 + z)^2)) \,,
\end{align}
where $ V$ is the scalar potential (\ref{eq-scalar-potential}) evaluated on the constant scalars $ \lambda_{1,2} = \Lambda_{1,2}$, and all the constants are given in terms of $ z$ as in (\ref{eq-constants-ansatz}). This combination should evaluate to $ 12L_0^2/L^2$ if one is to find the expected $ z$-independent 5D cosmological constant term with the correct coefficient relative to $ R$. Remarkably, setting
\begin{align}
    \mathpzc{a} & = \pm \frac{1}{4}
\end{align}
achieves precisely that, where the leftover sign choice will be important later on when we dimensionally reduce the BPS equations. Then the 7d action simplifies to
\begin{align}\label{eq:7dactionred5dApp}
    I_7 & = \frac{L_0^2}{16\pi G_7} \int \qty[ \star_5\qty(R + \frac{12}{L^2}) - \frac{1}{2}\qty(2\mathpzc{b}^2 + 2\mathpzc{b}\mathpzc{c} - \mathpzc{c}^2) F \wedge  \star_5 F + \mathpzc{b}^3 A \wedge F \wedge F]  \wedge \dd{\omega} \,.
\end{align}
To retain $ F = \dd{A}$ when the 5d equation of motion for $ F$ is derived we demand $ \qty(2\mathpzc{b}^2 + 2\mathpzc{b}\mathpzc{c} - \mathpzc{c}^2) = 3\mathpzc{b}^2$, which fixes $ \mathpzc{c} = \mathpzc{b}$. We integrate over $ \Sigma_{\mathfrak{g}}$: $ \int \dd{\omega} = 4\pi(\mathfrak{g} - 1)$, and rescale the last remaining constant to $ \mathpzc{b} = \mathpzc{x}/\sqrt{3}$. Then expressions (\ref{5DSUGRAaction}) and (\ref{eq-G7-G5-relation}) readily follow.

\subsection{Reducing the BPS equations}\label{sec-reducing-bps}

In this section we dimensionally reduce the 7d BPS equations (\ref{eq-7D-bps}) to the 5d BPS equation (\ref{eq-bps}). We introduce the following notation: 7d curved indices are labelled by $ M,N,\dots$, 5d curved indices are labelled by $ \mu,\nu,\dots$, 2d curved indices (on the Riemann surface) are labelled by $ m,n,\dots = x_1,x_2$, and flat indices in any dimension carry hats. The 7d metric decomposes as
\begin{align}
    g_{MN} & = e^{2f_0} \, g_{\mu\nu} \oplus e^{2g_0}L^{-2} \, g_{mn} \,,
\end{align}
where $ g_{\mu\nu}$ and $ g_{mn}$ are the metrics originating from $ \dd{s}^2_5$ and $ \dd{s}^2_2$ respectively (without the scaling factors). Then the 7D vielbeins decompose as
\begin{align}
    e\indices{_M^{\widehat{M}}} & = e^{f_0} \, e\indices{_\mu^{\widehat{\mu}}} \oplus e^{g_0} L \, e\indices{_m^{\widehat{m}}} \,.
\end{align}
From the reduction ansatz (\ref{eq-ansatz}) we see that the 7D gauge fields decompose as
\begin{align}
    \begin{aligned}
        S_{MNR} & = \mathpzc{c} \, e^{2f_0 + 2\Lambda_+} \, \frac{1}{2!} F^{\alpha\beta} \varepsilon_{\alpha\beta\mu\nu\rho}  \,, \\
        F_{1,2 MN} & = \mathpzc{b} \, e^{f_0 + 2\Lambda_{1,2}} \, F_{\mu\nu} \oplus \mathpzc{a} \, (1 \pm z)L \, \qty(\dd{\omega})_{mn}  \,,\\
        A_{1,2 M} & = \mathpzc{b} \, e^{f_0 + 2\Lambda_{1,2}} \, A_{\mu} \oplus \mathpzc{a} \, (1 \pm z)L \, \omega_{m}  \,.
    \end{aligned}
\end{align}
The 7d gamma matrices decompose as
\begin{align}
    \widetilde{\gamma}^{MN} = e^{-2f_0} \, \widetilde{\gamma}^{\mu\nu} \oplus e^{-2g_0}L^{-2} \, \widetilde{\gamma}^{mn} \,, \qquad \widetilde{\gamma}^{M} = e^{-f_0} \, \widetilde{\gamma}^{\mu} \oplus e^{-g_0}L^{-1} \, \widetilde{\gamma}^{m} \,.
\end{align}
The 7d spin connection decomposes trivially
\begin{align}
    \omega_{M \widehat{A} \widehat{B}} & = \omega_{\mu \widehat{\alpha} \widehat{\beta}} \oplus \omega_{m \widehat{a} \, \widehat{b}} \,,
\end{align}
and we note that the 2d spin connection has only one non-zero component given by
\begin{align}
    \omega_{x_1 \widehat{x}_1 \widehat{x}_2} & = - x_2^{-1} \,.
\end{align}
We also need the following relations between 7d gamma matrices
\begin{align}
    \widetilde{\gamma}^{MNR} & = \chi_7 \, \frac{1}{4!} \varepsilon^{MNRABCD} \widetilde{\gamma}_{ABCD} \,, \quad \widetilde{\gamma}^{MN} = \chi_7 \, \frac{1}{5!} \varepsilon^{MNABCDE} \widetilde{\gamma}_{ABCDE}
\end{align}
where $ \chi_7 = \pm 1$ dictates which of the two conventions one uses for the 1-index 7d gamma matrices, where the sign choice appears in the relation $ \widetilde{\gamma}^6 = \chi_7 \widetilde{\gamma}^0 \widetilde{\gamma}^{1} \dots \widetilde{\gamma}^{5}$. Using our metric Ansatz we obtain
\begin{align}
    \widetilde{\gamma}^{\mu\nu\rho} & = \chi_7 \, \frac{1}{2!} e^{-3f_0} \varepsilon^{\mu\nu\rho \alpha \beta} \widetilde{\gamma}_{\alpha \beta} \widetilde{\gamma}_{\widehat{x}_1 \widehat{x}_2} \,, \quad
    \widetilde{\gamma}^{\mu\nu} = \chi_7 \, \frac{1}{3!} e^{-2f_0 } \varepsilon^{\mu\nu \alpha \beta \gamma} \widetilde{\gamma}_{\alpha \beta \gamma} \widetilde{\gamma}_{\widehat{x}_1 \widehat{x}_2} \,.
\end{align}
Finally, we note that $ \dd{\omega}$, being the volume form on the Riemann surface, acts to lower 7d gamma matrix index, restricted to the Riemann surface, as
\begin{align}
    \widetilde{\gamma}^m \qty( \dd{\omega} )_{mn} & = - \widetilde{\gamma}_{n} \widetilde{\gamma}_{x_1 x_2} \,.
\end{align}
With this setup in mind, in (\ref{eq-7D-bps}) we set $ \mu = 2/L$ and take constant scalars: $ \lambda_{1,2} = \Lambda_{1,2}$. Then, the first dilatini variation reduces to
\begin{align}\label{eq:deltazeta1}
    \delta\widetilde{\zeta}_{1} & = \frac{L^{-1}}{2} \qty[\qty(e^{2\Lambda_{1}} - e^{-4\Lambda_+}) - \frac{\mathpzc{a}}{4} (1 + z) e^{-2g_0 - 2\Lambda_1} \, \widetilde{\gamma}_{\widehat{x}_1 \widehat{x}_2} \otimes \Gamma^{12}] \widetilde{\epsilon} \otimes \mathpzc{E} \nonumber \\
    & \qquad \qquad - \frac{e^{-f_0}}{16}  \widetilde{\gamma}^{\mu\nu} F_{\mu\nu} \qty[ \mathpzc{b} \, 1_{8} \otimes \Gamma^{12} + \chi_7 \, \mathpzc{c} \, \widetilde{\gamma}_{\widehat{x}_1 \widehat{x}_2} \otimes \Gamma^5 ] \widetilde{\epsilon} \otimes \mathpzc{E}\,. 
\end{align}
We find that the following projectors:
\begin{align}\label{eq-projectors}
    \qty(\widetilde{\gamma}_{\widehat{x}_1 \widehat{x}_2} \otimes 1_{4}) (\widetilde{\epsilon} \otimes \mathpzc{E})  = p_2 \, \qty(1_{8} \otimes \Gamma^{1 2}) (\widetilde{\epsilon} \otimes \mathpzc{E}) \,, \qquad \qty(1_{8} \otimes \Gamma^5) \qty(\widetilde{\epsilon} \otimes \mathpzc{E}) = p_5 \, \widetilde{\epsilon} \otimes \mathpzc{E} \,,
\end{align}
together with the identities $ \qty(\widetilde{\gamma}_{\widehat{x}_1 \widehat{x}_2})^2 = -1_{8}$ and $ \qty(\Gamma^{1 2})^2 = -1_{4}$, imply that the two brackets in \eqref{eq:deltazeta1} vanish when:
\begin{align}\label{eq-conditions}
    p_2 & = s_{a} \,, \quad p_5 = - \chi_7 \, s_a \,; \qquad \mathpzc{a} = \frac{s_a}{4} \,, \quad \mathpzc{c} = \mathpzc{b} = \frac{\mathpzc{x}}{\sqrt{3}}\,,
\end{align}
where $ s_a = \pm 1$ and the latter two conditions come from consistency of the reduced action as discussed around \eqref{eq:7dactionred5dApp}. The second dilatini variation automatically vanishes upon using (\ref{eq-projectors}) and (\ref{eq-conditions}) giving no new constraints. Next, we manipulate the gravitini variation when $ M = m$, keeping in mind that $ \qty(1_{8} \otimes \Gamma^5) \qty(\widetilde{\epsilon} \otimes \mathpzc{E}) = p_5 \, \widetilde{\epsilon} \otimes \mathpzc{E} \implies \qty(1_{8} \otimes \Gamma^{34}) \qty(\widetilde{\epsilon} \otimes \mathpzc{E}) = p_5 \qty(1_{8} \otimes \Gamma^{1 2}) \qty(\widetilde{\epsilon} \otimes \mathpzc{E})$:
\begin{align}\nonumber
    \delta \widetilde{\psi}_m & \overset{p_5 = 1}{=} \Bigg[ \partial_m - \frac{1}{2} x_2^{-1} \delta_{mx_1} \qty(\widetilde{\gamma}_{\widehat{x}_1 \widehat{x}_2} \otimes 1_{4} - 4\mathpzc{a} \, 1_{8} \otimes \Gamma^{12}) \\
    & \qquad \qquad + \frac{1}{2} e^{-f_0 + g_0} \widetilde{\gamma}_m \qty( 1_{8} \otimes 1_{4} + 4\mathpzc{a} \, \widetilde{\gamma}_{\widehat{x}_1 \widehat{x}_2} \otimes \Gamma^{12} ) \Bigg]\widetilde{\epsilon} \otimes \mathpzc{E} \,,
\end{align}
where we have set $ p_5 = 1$, since otherwise the terms in the bracket are $ z$-dependent, and no projector can make them vanish. Upon using (\ref{eq-projectors}) and (\ref{eq-conditions}), with $ p_5 = 1$, the two brackets above vanish. For the whole equation to vanish we find that the spinor should be independent of the Riemann surface coordinates. We thus find the following constraints between the constants defined above
\begin{align}\label{eq-conditions2}
    p_2 & = -\chi_7 \,, \quad p_5 = 1 \,; \qquad \mathpzc{a} = - \frac{\chi_7}{4} \,, \quad \mathpzc{c} = \mathpzc{b} = \frac{\mathpzc{x}}{\sqrt{3}} \,;  \qquad \partial_{m} \qty(\widetilde{\epsilon} \otimes \mathpzc{E}) = 0 \,.
\end{align}
Finally, we manipulate the gravitini variation for $ M = \mu$ to find: 
\begin{align}
    \delta \widetilde{\psi}_\mu & = \Bigg( \bigg[ \partial_\mu + \frac{1}{4} \omega_{\mu \widehat{\alpha} \widehat{\beta}} \widetilde{\gamma}^{\widehat{\alpha} \widehat{\beta}} + \chi_7 \frac{\mathpzc{x}}{8 \sqrt{3}} \qty( \widetilde{\gamma}_{\mu\alpha\beta} F^{\alpha\beta} -  4 \, F_{\mu\alpha} \widetilde{\gamma}^\alpha  ) \widetilde{\gamma}_{\widehat{x}_1 \widehat{x}_2} \nonumber \\
    & \qquad \qquad + \frac{1}{2L} \qty( \widetilde{\gamma}_\mu - \chi_7 \, \sqrt{3} \, \mathpzc{x} \, A_\mu \widetilde{\gamma}_{\widehat{x}_1 \widehat{x}_2} ) \bigg] \widetilde{\epsilon} \Bigg) \otimes \mathpzc{E} \,, \nonumber
\end{align}
where we have already imposed (\ref{eq-projectors}) and (\ref{eq-conditions2}). To complete the reduction we further split the 7d gamma matrices and spinor as
\begin{align}\label{eq-splitting}
    \widetilde{\gamma}_{\mu} & = \gamma_{\mu} \otimes g_1 \,, \quad \widetilde{\gamma}_{\widehat{x}_1 \widehat{x}_2} = 1_{4} \otimes g_2 \,, \quad \widetilde{\epsilon} = \epsilon  \otimes \eta \,,
\end{align}
where $ \gamma_\mu$ and $ \epsilon$ are 5d gamma matrices and spinor, $ \eta$ is a 2d spinor and $ (g_1, \, g_2)$ are (yet undetermined) matrices in the 2d Clifford algebra. Then the round bracket in the gravitini variation above becomes
\begin{align}
    & \bigg[ \partial_\mu + \frac{1}{4} \omega_{\mu \widehat{\alpha} \widehat{\beta}} \gamma^{\widehat{\alpha} \widehat{\beta}} \otimes g_1^2 + \chi_7 \frac{\mathpzc{x}}{8 \sqrt{3}} \qty( {\gamma}_{\mu\alpha\beta} F^{\alpha\beta} \otimes g_1^3g_2 -  4 \, F_{\mu\alpha} \gamma^\alpha \otimes g_1g_2 ) \nonumber \\
    & \quad - \frac{1}{2L} \qty( - \gamma_\mu \otimes g_1 + \chi_7 \, \sqrt{3} \, \mathpzc{x} \, A_\mu \, 1_{4} \otimes g_2 ) \bigg] \epsilon \otimes \eta \,. \nonumber
\end{align}
This equation reduces to (\ref{eq-bps}) if we are able to find matrices $ (g_1, \, g_2)$ in the 2d Clifford algebra that satisfy
\begin{align}\label{eq-conditions3}
    g_1 \eta = - \eta \,, \quad g_2 \eta = \iu \, \chi_7 \, \eta \,.
\end{align}
Upon using (\ref{eq-conditions2}), we find consistency of (\ref{eq-conditions3}) with the projectors (\ref{eq-projectors}) if
\begin{align}
    \Gamma^{1 2} \mathpzc{E} & = - \iu \, \mathpzc{E} \,, \quad \Gamma^5 \mathpzc{E} = \mathpzc{E} \,:  \quad \implies \quad \mathpzc{E}_1 = \mathpzc{E}_2 = \mathpzc{E}_3 = 0 \,.
\end{align}
In turn, upon using the projectors again, the latter condition demands $ \eta \propto (1, \, c_7)^T$. Thus, we fix the matrices $ (g_1, \, g_2)$ as
\begin{align}
    g_1 & = - \chi_7 \, \sigma_1 \,, \quad g_2 = \iu \, \sigma_1 \,.
\end{align}
This completes the reduction of the 7d supersymmetry variations to the 5d ones.

\section{Details on the analysis of the 5d CCLP solution}\label{sec-CCLP-appendix}

Here we collect some calculational details related to the analysis of the 5d CCLP solution discussed in Section~\ref{sec-thermo-CCLP}.

\subsection{Derivation of the Euclidean 5d action}
We define the $ \varepsilon$-pseudo-tensor as
\begin{align}
    \varepsilon_{01234} & = \sqrt{-g} \,, \quad \varepsilon^{01234} = \frac{s}{\sqrt{-g}} \,,
\end{align}
where $ s = -1$ denotes the signature of the Lorentzian spacetime and $ g = \text{det}(g_{\mu\nu})$ (not to be confused with the inverse AdS scale parameter $ g$ used in Section \ref{sec-thermo-CCLP}). In our conventions, described in Appendix \ref{sec-conventions}, we write the action (\ref{5DSUGRAaction}) in component form as
\begin{align}\label{eq-action-lor-comp}
    I & = \frac{1}{16\pi G_5} \int \dd[5]{x} \sqrt{-g} \qty( R + \frac{12}{L^2} - \frac{\mathpzc{x}^2}{4} F^{\mu\nu}F_{\mu\nu} + \frac{s \mathpzc{x}^3}{12\sqrt{3}} \varepsilon^{\nu\alpha\beta\gamma\delta} A_\nu F_{\alpha\beta}F_{\gamma\delta}) \,,
\end{align}
The Euclidean theory is obtained from the Lorentzian one by performing Wick rotation
\begin{align}\label{eq-wick}
    t & \to - \iu\, \tau \,.
\end{align}
Under Wick rotation the components of a generic tensor transform as
\begin{align}
    T_{0ij\dots} & \to  \iu \, T_{0ij\dots} \,, \quad T^{0ij\dots} \to  -\iu \, T^{0ij\dots} \,,
\end{align}
where $ i,j,\dots = 1,2,\dots$ label coordinates distinct from $ t$ or $ \tau$, in the respective coordinates systems. In particular, we have
\begin{align}
    \dd[5]{x} & \to -\iu \dd[5]{x} \,, \quad \sqrt{-g} \to \sqrt{g} \,, \quad R \to R \,, \quad F^{\mu\nu}F_{\mu\nu} \to F^{\mu\nu} F_{\mu\nu} \,.
\end{align}
The components of a pseudo-tensor transform involving also the sign of the coordinate transformation. In particular, the $ \varepsilon$-pseudo-tensor transforms as
\begin{align}
    \varepsilon_{0ijkl} & \to (-\iu) \cdot \iu \, \varepsilon_{0ijkl} = \varepsilon_{0ijkl} \,, \quad \varepsilon^{0ijkl} \to \iu \cdot (-\iu) \cdot s \, \varepsilon^{0ijkl} = s \, \varepsilon^{0ijkl} \,,
\end{align}
where we have included a factor of $ s = -1$ in the upper components to ensure that the Euclidean $ \varepsilon$-pseudo-tensor is adapted to the positive Euclidean signature
\begin{align}
    \varepsilon_{01234} & = \sqrt{g} \,, \quad \varepsilon^{01234} = \frac{1}{\sqrt{g}} \,.
\end{align}
Thus, the Chern-Simons term transforms as
\begin{align}
    \varepsilon^{\nu\alpha\beta\gamma\delta} A_{\nu} F_{\alpha\beta} F_{\gamma\delta} & = \qty(\varepsilon^{0ijkl} A_0 F_{ij} F_{kl} - \varepsilon^{0ijkl} A_i F_{0j}F_{kl} + \dots) \nonumber \\
    & \to \iu\, s \qty(\varepsilon^{0ijkl} A_{0} F_{ij} F_{kl} - \varepsilon^{0ijkl} A_{i} F_{0j}F_{kl} + \dots) \nonumber \\
    & = \iu\, s\, \varepsilon^{\nu\alpha\beta\gamma\delta} A_{\nu} F_{\alpha\beta} F_{\gamma\delta}\,,
\end{align}
and the Euclidean action takes the form
\begin{align}\label{eq-action-eucl}
    I_{\mathpzc{E}} & = - \frac{1}{16\pi G_5} \int \dd[5]{x} \sqrt{g}\qty( R + \frac{12}{L^2} - \frac{\mathpzc{x}^2}{4} F^{\mu\nu} F_{\mu\nu} + \frac{\iu\, \mathpzc{x}^3}{12\sqrt{3}} \varepsilon^{\nu\alpha\beta\gamma\delta} A_{\nu} F_{\alpha\beta} F_{\gamma\delta}) \,,
\end{align}
Using this action one can derive the following Euclidean equations of motion
\begin{align}\label{eq-eom-eucl}
    \begin{aligned}
        R_{\mu\nu} + \frac{4}{L^2} g_{\mu\nu} - \frac{\mathpzc{x}^2}{2}F\indices{_\mu^\alpha} F_{\nu\alpha} + \frac{\mathpzc{x}^2}{12}g_{\mu\nu} F^{\alpha\beta} F_{\alpha\beta} = 0 \,, \\
        \nabla_\mu F^{\mu\nu} + \frac{\iu\, \mathpzc{x}}{4\sqrt{3}} \varepsilon^{\nu\alpha\beta\gamma\delta} F_{\alpha\beta} F_{\gamma\delta} = 0 \,.
    \end{aligned}
\end{align}

\subsection{Near horizon Lorentzian geometry}
In the near horizon region we first perform the coordinate transformation 
\begin{align}
    r & = r_+ + \varrho^2 \,,
\end{align}
then we expand the Lorentzian metric (\ref{eq-cclp-hopf1}) for small $ \varrho$ to obtain
\begin{align}\label{eq-near-hor-met}
    \dd{s}^2_{\text{CCLP},\mathpzc{H}} & = \widetilde{g}_{\varrho\varrho} \qty(\dd{\varrho}^2 -  \qty(\frac{2\pi}{\beta^{\mathpzc{H}}})^2 \varrho^2 \dd{t}^2) + \widetilde{g}_{\eta\eta} \dd{\eta}^2 + 2\, \widetilde{g}_{12} \qty( \dd{\xi_1} - \Omega_1^{\mathpzc{H}} \dd{t}) \qty(\dd{\xi_2} - \Omega_2^{\mathpzc{H}} \dd{t}) \nonumber\\
    & \qquad + \widetilde{g}_{11} \qty( \dd{\xi_1} - \Omega_1^{\mathpzc{H}} \dd{t} )^2 + \widetilde{g}_{22} \qty( \dd{\xi_2} - \Omega_2^{\mathpzc{H}} \dd{t} )^2 \,.
\end{align}
Here we defined the constants
\begin{align}\label{eq-hor-equals-pot}
    \beta^{\mathpzc{H}} & = \frac{2\pi r_+\qty[(r_+^2 + a^2)(r_+^2 + b^2) + abq]}{r_+^4\qty[1 + g^2(2r_+^2 + a^2 + b^2)] - (ab + q)^2} \,,\nonumber\\
    \Omega_1^{\mathpzc{H}} & = \frac{a(r_+^2 + b^2)(1 + g^2r_+^2) + bq}{(r_+^2 + a^2)(r_+^2 + b^2) + abq} \,, \quad \Omega_2^{\mathpzc{H}} = \eval{\Omega_1^{\mathpzc{H}}}_{a \leftrightarrow b} \,,
\end{align}
and $ (\widetilde{g}_{\varrho\varrho},\, \widetilde{g}_{\eta\eta},\, \widetilde{g}_{11},\, \widetilde{g}_{22},\, \widetilde{g}_{12})$ are functions of $ (\varrho,\eta)$ expanded to the relevant orders in the small variable $ \varrho$:
\begin{align}
    \begin{aligned}
        \widetilde{g}_{\varrho \varrho} & = \frac{4 \rho_+^2}{\Delta_{r,+}'} \,, \qquad \widetilde{g}_{\eta\eta} = \frac{\rho_+^2}{\Delta_\eta} \,, \\
        \frac{\widetilde{g}_{11}}{\mathpzc{S}_\eta^2} & = - \frac{(\Delta_\eta - \Xi_a)\qty(a^2 f_+  + 2abq\Xi_a \rho_+^2) - (a^2 + r_+^2)\Xi_a(\Xi_a - \Xi_b)\rho_+^4}{\Xi_a^2(\Xi_a - \Xi_b)\rho_+^4}  \\
        & \quad - \frac{(\Delta_\eta - \Xi_a)\qty[a^2f_+'\rho_+^2 - 2a\qty(a f_+ + bq \Xi_a \rho_+^2)\qty(\rho_+^2)'] - 2r_+ \Xi_a(\Xi_a - \Xi_b)\rho_+^6}{\Xi_a^2(\Xi_a - \Xi_b)\rho_+^6} \varrho^2\,, \\
        \frac{\widetilde{g}_{12}}{\mathpzc{S}_\eta^2 \mathpzc{C}_\eta^2} & = \frac{ab f_+ + (b^2\Xi_a + a^2 \Xi_b) q\rho_+^2}{\Xi_a \Xi_b \rho_+^4} \\
        & \quad + \frac{ab\qty[f_+' \rho_+^2 - 2 f_+ \qty(\rho_+^2)'] - (b^2 \Xi_a + a^2 \Xi_b)\rho_+^2 \qty(\rho_+^2)'}{\Xi_a \Xi_b \rho_+^6} \varrho^2\,, \\
        \frac{\widetilde{g}_{22}}{\mathpzc{C}_\eta^2} & = \frac{(\Delta_\eta - \Xi_b)\qty(b^2 f_+  + 2abq\Xi_b \rho_+^2) + (b^2 + r_+^2)\Xi_b(\Xi_a - \Xi_b)\rho_+^4}{\Xi_b^2(\Xi_a - \Xi_b)\rho_+^4}  \\
        & \quad + \frac{(\Delta_\eta - \Xi_b)\qty[b^2f_+'\rho_+^2 - 2b\qty(b f_+ + aq \Xi_b \rho_+^2)\qty(\rho_+^2)'] + 2r_+ \Xi_b(\Xi_a - \Xi_b)\rho_+^6}{\Xi_b^2(\Xi_a - \Xi_b)\rho_+^6} \varrho^2 \,,
    \end{aligned}
\end{align}
where we have defined the shorthands
\begin{align}
    f_+ & = f(r_+,\eta) \,, \quad \rho^2_+ = \rho^2(r_+,\eta) \,, \quad \mathpzc{S}_\eta = \sin\eta \,, \quad \mathpzc{C}_\eta = \cos\eta \,, \nonumber \\
    f_+' & = \pdv{f}{r}\qty(r_+,\eta) \,, \quad \qty(\rho_+^2)' = \pdv{\rho^2}{r}\qty(r_+,\eta) \,, \quad \Delta_{r,+}' = \pdv{\Delta_r}{r}\qty(r_+) \,. \nonumber
\end{align}

\subsection{Asymptotic Lorentzian geometry}
To obtain the Lorentzian metric in the asymptotic region we perform the coordinate transformation
\begin{align}
    r & = \widehat{r}\qty(1 + \frac{1}{\widehat{r}^2}f_1(\widehat{\eta}) + \frac{1}{\widehat{r}^4}f_2(\widehat{\eta})) \,, \quad \eta = \widehat{\eta} + \frac{1}{\widehat{r}^4}h_1(\widehat{\eta}) + \frac{1}{\widehat{r}^6}h_2(\widehat{\eta}) \,,
\end{align}
where ($ f_{1,2}$, $ h_{1,2}$) are functions of $ \widehat{\eta}$ that will be fixed by demanding that the asymptotic metric takes the Fefferman-Graham form. We expand the components of the transformed metric to (and including) the following orders:
\begin{align}\nonumber
    \mqty{
        \widehat{g}_{t t} & \widehat{g}_{t1} & \widehat{g}_{t2} & \widehat{g}_{\widehat{r}\widehat{r}} & \widehat{g}_{\widehat{r} \widehat{\eta}} & \widehat{g}_{\widehat{\eta} \widehat{\eta}} & \widehat{g}_{11} & \widehat{g}_{12} & \widehat{g}_{22} \\
        O\qty(\widehat{r}^{-2}) & O\qty(\widehat{r}^{-2}) & O\qty(\widehat{r}^{-2}) & O\qty(\widehat{r}^{-6}) & O\qty(\widehat{r}^{-5}) & O\qty(\widehat{r}^{-2}) & O\qty(\widehat{r}^{-2}) & O\qty(\widehat{r}^{-2}) & O\qty(\widehat{r}^{-2}) \,.
    }
\end{align}
Putting the metric in Fefferman-Graham form amounts to having $ \widehat{g}_{\widehat{r}\widehat{r}} = g^{-2} \widehat{r}^{-2}$ and $ \widehat{g}_{\widehat{r} \widehat{\eta}} = 0$. Demanding that the $ O\qty(\widehat{r}^{-4})$ and $ O\qty(\widehat{r}^{-6})$ terms of $ \widehat{g}_{\widehat{r}\widehat{r}}$ and the $ O\qty(\widehat{r}^{-3})$ and $ O\qty(\widehat{r}^{-5})$ terms of $ \widehat{g}_{\widehat{r}\widehat{\eta}}$ vanish achieves that and uniquely fixes
\begin{align}\nonumber
    \begin{aligned}
        f_1(\widehat{\eta}) & = -\frac{1}{4g^2} \qty[2 - \Xi_a - \Xi_b + \Delta_\eta(\widehat{\eta}) ] \,, \\
        f_2(\widehat{\eta}) & = \frac{1}{8 g^4}\qty[1 - 2g^2m - \Xi_a \sin^2\widehat{\eta} - \Xi_b \cos^2\widehat{\eta} - \cos^2\widehat{\eta} \sin^2\widehat{\eta} \qty(\Xi_a - \Xi_b)^2] \,, \\
        h_1(\widehat{\eta}) & = -\frac{\Xi_a - \Xi_b}{8g^4}\cos\widehat{\eta} \sin^2 \widehat{\eta} \, \Delta_\eta(\widehat{\eta}) \,, \\
        h_2(\widehat{\eta}) & = - \frac{\Xi_a - \Xi_b}{16g^6}\cos \widehat{\eta} \sin^2 \widehat{\eta} \qty[2 \Xi_a^2\cos^2\widehat{\eta} - \Xi_a \Xi_b + 2\Xi_b^2 \sin^2\widehat{\eta} - 3\cos^2\widehat{\eta} \sin^2\widehat{\eta}(\Xi_a - \Xi_b)^2] \,.
    \end{aligned}
\end{align}
The resulting Fefferman-Graham asymptotic Lorentzian metric reads
\begin{align}\label{eq-cclp-asymp}
    \dd{s}^2_{\text{CCLP},\mathpzc{A}} & = \frac{1}{g^2 \widehat{r}^2} \dd{\widehat{r}}^2 + \widehat{g}_{tt} \dd{t}^2 + 2\, \widehat{g}_{t1} \dd{t} \dd{\xi_1} + 2\, \widehat{g}_{t2} \dd{t} \dd{\xi_2} + \widehat{g}_{\widehat{\eta}\widehat{\eta}} \dd{\widehat{\eta}}^2 \nonumber\\
    & \qquad +  \widehat{g}_{11} \dd{\xi_1}^2 + 2 \, \widehat{g}_{12} \dd{\xi_1} \dd{\xi_2} + \widehat{g}_{22} \dd{\xi_2}^2 \,.
\end{align}
In the equation above $ (\widehat{g}_{tt}, \, \widehat{g}_{t1}, \, \widehat{g}_{t2}, \, \widehat{g}_{\widehat{\eta}\widehat{\eta}}, \, \widehat{g}_{11}, \, \widehat{g}_{12}, \, g_{22})$ are function of $ (\widehat{r},\widehat{\eta})$, expanded to the relevant orders in the large variable $ \widehat{r}$: 
\begin{align}
    \begin{aligned}
        \widehat{g}_{tt}&=-\frac{g^2 \widehat{\Delta}_\eta \, \widehat{r}^2 }{\Xi_a \Xi_b}-\frac{\widehat{\Delta}_\eta\left(\Xi_a+\Xi_b-\widehat{\Delta}_\eta\right)}{2 \Xi_a \Xi_b} \\
        & \quad +\frac{2 \widehat{\Delta}_\eta^2\left(a b g^2 q+m\right)-\frac{\Xi_a \Xi_b \widehat{\Delta}_\eta\left(-2 \widehat{\Delta}_\eta\left(\Xi_a+\Xi_b\right)+\left(\Xi_a+\Xi_b\right)^2+\widehat{\Delta}_\eta^2+8 g^2 m\right)}{16 g^2}}{\Xi_a^2 \Xi_b^2 \, \widehat{r}^2} \,,\\
        \frac{\widehat{g}_{t1}}{\widehat{\mathpzc{S}}^2_\eta}&=-\frac{\widehat{\Delta}_\eta \left(a^2 b g^2 q+2 a m+b q\right)}{\Xi_a^2 \Xi_b \, \widehat{r}^2} \,,\\
        \frac{\widehat{g}_{t 2}}{\widehat{\mathpzc{C}}^2_\eta}&=-\frac{\widehat{\Delta}_\eta \left(a b^2 g^2 q+a q+2 b m\right)}{\Xi_a \Xi_b^2 \, \widehat{r}^2} \,,\\
        \widehat{g}_{\widehat{\eta} \widehat{\eta}}&= \frac{\widehat{r}^2}{\widehat{\Delta}_\eta} + \frac{\Xi_a+\Xi_b-3 \widehat{\Delta}_\eta}{2 g^2 \widehat{\Delta}_\eta}+\frac{-6 \widehat{\Delta}_\eta\left(\Xi_a+\Xi_b\right)+\left(\Xi_a+\Xi_b\right)^2+9 \widehat{\Delta}_\eta^2+8 g^2 m}{16 g^4 \widehat{\Delta}_\eta \, \widehat{r}^2} \,,\\
        \frac{\widehat{g}_{11}}{\widehat{\mathpzc{S}}^2_\eta}&=\frac{\widehat{r}^2}{\Xi_a}-\frac{\Xi_a-\Xi_b+\widehat{\Delta}_\eta}{2 g^2 \Xi_a} \\
        & \quad -\frac{\frac{8 g^2\left(4 a^2 g^2 m \widehat{\Delta}_\eta+4 a b g^2 q\left(\widehat{\Delta}_\eta-\Xi_a\right)+m \Xi_a\left(3 \Xi_a+\Xi_b-4\right)\right)}{\Xi_a-\Xi_b}-\Xi_a\left(\Xi_a-\Xi_b+\widehat{\Delta}_\eta\right)^2}{16 g^4 \Xi_a^2 \, \widehat{r}^2} \,,\\
        \frac{\widehat{g}_{12}}{\widehat{\mathpzc{S}}^2_\eta \widehat{\mathpzc{C}}^2_\eta}&= \frac{a^2 q+2 a b m+b^2 q}{\Xi_a \Xi_b \, \widehat{r}^2} \,,\\
        \frac{\widehat{g}_{22}}{\widehat{\mathpzc{C}}^2_\eta}&= \frac{\widehat{r}^2 }{\Xi_b}+\frac{\Xi_a-\Xi_b-\widehat{\Delta}_\eta}{2 g^2 \Xi_b}\\
        & \quad + \frac{\frac{8 g^2\left(4 b^2 g^2 m \widehat{\Delta}_\eta + 4 a b g^2 q\left(\widehat{\Delta}_\eta-\Xi_b\right)+m \Xi_b\left(\Xi_a+3 \Xi_b-4\right)\right)}{\Xi_a-\Xi_b}+\Xi_b\left(-\Xi_a+\Xi_b+\widehat{\Delta}_\eta\right)^2}{16 g^4 \Xi_b^2 \, \widehat{r}^2} \,,
        \end{aligned}
\end{align}
where we have defined the shorthands
\begin{align}\nonumber
    \widehat{\Delta}_{\eta} & \equiv \Delta_\eta(\widehat{\eta}) \,, \quad \widehat{\mathpzc{S}}_\eta = \sin\widehat{\eta} \,, \quad \widehat{\mathpzc{C}}_\eta = \cos\widehat{\eta}\,.
\end{align}

\subsection{Potentials: $ (\Omega_1, \, \Omega_2, \, \beta, \, \Phi)$}\label{sec-potentials}
The thermodynamic potentials associated with the two angular momentum parameters of the CCLP black hole are defined as
\begin{align}
    \Omega_i & = \Omega_i^{\mathcal{H}} - \Omega_i^{\mathcal{B}} \,,  \quad \Omega_i^{\mathcal{H}} = \eval{\dv{\xi_i}{t}}_{r \to r_+} \,, \quad \Omega_i^{\mathcal{B}} = \eval{\dv{\xi_i}{t}}_{r \to \infty} \,, \quad i = 1,2 \,.
\end{align}
From (\ref{eq-near-hor-met}) and (\ref{eq-conf-bdy}) it is easy to see that $ \Omega_i^{\mathcal{H}} = \Omega_i^{\mathpzc{H}}$ and $ \Omega_i^{\mathcal{B}} = 0$, or explicitly
\begin{align}
    \Omega_1 & = \frac{a(r_+^2 + b^2)(1 + g^2r_+^2) + bq}{(r_+^2 + a^2)(r_+^2 + b^2) + abq} \,, \quad \Omega_2 = \frac{b(r_+^2 + a^2)(1 + g^2r_+^2) + aq}{(r_+^2 + a^2)(r_+^2 + b^2) + abq} \,.
\end{align}
The inverse temperature of the CCLP black hole can be obtained from the Wick rotated near horizon metric
\begin{align}\label{eq-near-hor-eucl-met}
    \dd{s}^2_{\text{CCLP},\mathpzc{H},\mathpzc{E}} & = \widetilde{g}_{\varrho\varrho} \qty(\dd{\varrho}^2 + \qty(\frac{2\pi}{\beta^{\mathpzc{H}}})^2 \varrho^2 \dd{\tau}^2) + \widetilde{g}_{\eta\eta} \dd{\eta}^2 + 2\, \widetilde{g}_{12} \qty( \dd{\xi_1} + \iu \, \Omega_1 \dd{t}) \qty(\dd{\xi_2} + \iu \, \Omega_2 \dd{t}) \nonumber\\
    & \qquad + \widetilde{g}_{11} \qty( \dd{\xi_1} + \iu \, \Omega_1 \dd{t} )^2 + \widetilde{g}_{22} \qty( \dd{\xi_2} + \iu \, \Omega_2 \dd{t} )^2 \,,
\end{align}
where we have already implemented the relations $ \Omega_i^{\mathpzc{H}} = \Omega_i$. The above metric is a warped fibration of $ S^3$ over $ \mathbb{R}^2$. Demanding absence of conical singularity at the origin of the $ \mathbb{R}^2$ fixes
\begin{align}
    \tau \sim \tau + \beta^{\mathpzc{H}} \,.
\end{align}
Demanding that the $ S^3$ fibration is also well defined at the origin of the $ \mathbb{R}^2$ shows that going around the Euclidean time circle once should be accompanied by the identifications
\begin{align}\label{eq-twisted-identification-eucl}
    \qty(\tau,\, \xi_1,\, \xi_2) & \sim \qty(\tau + \beta^{\mathpzc{H}},\, \xi_1 - \iu\, \Omega_1 \beta^{\mathpzc{H}},\, \xi_2 - \iu\, \Omega_2 \beta^{\mathpzc{H}}) \,.
\end{align}
From (\ref{eq-conf-bdy}) we see that the Wick rotated conformal boundary metric is regular for any length of the Euclidean time circle (also regular for non-compact $ \tau$ for that matter). For consistency, we translate the IR imposed periodicity on $ \tau$ to the boundary. Back in the Lorentzian picture, the periodicity in the Euclidean time has the interpretation of inverse temperature so we define $ \beta = \beta^{\mathpzc{H}}$, or explicitly
\begin{align}\label{eq-temperature}
    \beta & = \frac{2\pi r_+\qty[(r_+^2 + a^2)(r_+^2 + b^2) + abq]}{r_+^4\qty[1 + g^2(2r_+^2 + a^2 + b^2)] - (ab + q)^2} \,.
\end{align}
The inverse temperature can also be obtained directly from the Lorentzian description as follows. First, note the surface $ r = r_+$ is an event horizon as it is generated by a Killing vector
\begin{align}\label{eq-killing}
    V & = V^\mu\partial_\mu = \pdv{}{t} + \Omega_1 \pdv{}{\xi_1} + \Omega_2 \pdv{}{\xi_2}; \quad \nabla_\mu V_\nu + \nabla_\nu V_\mu = 0 \,,
\end{align}
that becomes null on that surface
\begin{align}
    \eval{V_\mu V^\mu}_{r = r_+} & = 0 \,.
\end{align}
Note that contrary to the familiar similar expressions from spacetimes with flat asymptotics, $ V_\mu V^\mu$ is not normalized to unity at infinity. Using this Killing vector we calculate the surface gravity on the horizon from
\begin{align}
    \kappa_{\mathpzc{H}} & = \eval{\sqrt{- \frac{1}{2} \nabla^\mu V^\nu \nabla_\mu V_\nu}}_{r \to r_+} \,.
\end{align}
The inverse temperature is then given by $ \beta = 2\pi/\kappa_{\mathpzc{H}}$. Simplifying this expression we obtain precisely (\ref{eq-temperature}). The electrostatic potential is defined as
\begin{align}
    \Phi & = \eval{V^\mu A_\mu}_{r = r_+} - \eval{V^\mu A_\mu}_{r \to \infty} \,.
\end{align}
It is instructive to look at the above terms separately:
\begin{align}
    \eval{V^\mu A_\mu}_{r = r_+} & = \frac{\sqrt{3} q r_+^2}{\mathpzc{x}\qty[(r_+^2 + a^2)(r_+^2 + b^2) + abq]} + \alpha \,, \qquad \eval{V^\mu A_\mu}_{r \to \infty} = \alpha \,.
\end{align}
We see that the definition of $ \Phi$ given above is independent of the choice of gauge parameter $ \alpha$. The final expression reads
\begin{align}\label{eq-el-pot}
    \Phi & = \frac{\sqrt{3} q r_+^2}{\mathpzc{x}\qty[(r_+^2 + a^2)(r_+^2 + b^2) + abq]} \,.
\end{align}
For consistency, one needs to ensure that $ A^\mu A_\mu$ is regular on the horizon. However, $ \eval{A^\mu A_\mu}_{r = r_+}$ diverges as $ O\qty( (r - r_+)^{-1} )$. Demanding that the coefficient of the divergent term vanishes fixes the gauge parameter $ \alpha$ to
\begin{align}\label{eq-alpha-to-Phi}
    \alpha & = -\Phi \,.
\end{align}
With this at hand we find that the following relations hold
\begin{align}\label{eq-reg-A}
    \eval{A^\mu A_\mu}_{r = r_+} & = 0 \,, \quad \eval{V^\mu A_\mu}_{r = r_+} = 0 \,.
\end{align}

\subsection{Charges: $ (J_1, \, J_2, \, Q)$, via boundary integrals}\label{sec-charges1}

To obtain the two conserved angular momenta $ (J_1,J_2)$ we first note that the boundary metric (\ref{eq-conf-bdy}) has three independent Killing vectors
\begin{align}
    K_{(t)} & = \pdv{}{t} \,, \quad K_{(\xi_1)} = \pdv{}{\xi_1} \,, \quad K_{(\xi_2)} = \pdv{}{\xi_2} \,.
\end{align}
To each of them we associate a Killing form as
\begin{align}
    \widetilde{K}_{( \cdot )} & = K^\mu_{( \cdot )} g_{\mu\nu} \dd{x^\nu} \,.
\end{align}
Then, the angular momenta of the black hole are defined via the following Komar integrals
\begin{align}
    J_i & = \lim_{r \to \infty} \frac{1}{16 \pi G_5}\int_{S^3} \star_5 \dd{\widetilde{K}_{(\xi_i)}} \,, \quad i = 1,2 \,.
\end{align}
The final results read
\begin{align}\label{eq-ang-mom}
    J_1 & = \frac{\pi\qty[2am + bq(1 + a^2g^2)]}{4\Xi_a^2 \Xi_b G_5} \,, \quad J_2 = \frac{\pi\qty[2bm + aq(1 + b^2g^2)]}{4\Xi_a \Xi_b^2 G_5} \,.
\end{align}
To calculate the electric charge $ Q$ one needs a conserved current arising from the Maxwell equation. The following manipulation identifies such current
\begin{align}
    0 & = \sigma\qty[\dd{} \star_5 F - \frac{1}{\sqrt{3}} F \wedge F] = \sigma\qty[ \dd{} \star_5 F - \frac{1}{\sqrt{3}} \dd{\qty(F \wedge A)}] = \dd{ \qty[\sigma\qty(\star_5 F - \frac{1}{\sqrt{3}} F \wedge A)]} \,, \nonumber
\end{align}
where we have included an arbitrary multiplicative parameter $ \sigma$. The asymptotic electric charge is then given by
\begin{align}
    Q & = \lim_{r \to \infty} \frac{\sigma}{16\pi G_5} \int_{S^3} \qty( \star_5 F - \frac{1}{\sqrt{3}} F \wedge A ) = - \frac{\sigma \sqrt{3} \pi q}{4\mathpzc{x} \Xi_a \Xi_b G_5} \,.
\end{align}
Consistency with the quantum statistical relation (\ref{eq-qsr}) fixes $ \sigma = -\mathpzc{x}^2$. Thus, the electric charge is given by
\begin{align}\label{eq-el-charge}
    Q & = \frac{\mathpzc{x} \sqrt{3} \pi q}{4 \Xi_a \Xi_b G_5} \,.
\end{align}

\subsection{Euclidean on-shell action}\label{sec-on-shell-action}
The on-shell action can be calculated using holographic renormalization. In the Lorentzian setting one begins by augmenting the action (\ref{eq-action-lor-comp}) with the following boundary terms
\begin{align}\label{eq-bdy-terms-hren}
    \begin{aligned}
        I_{\text{GH}} & = \frac{1}{8\pi G_5} \int \dd[4]{x} \sqrt{-\gamma} \, K \,, \\
        I_{\text{ct}} & = \frac{1}{8\pi G_5} \int \dd[4]{x} \sqrt{-\gamma} \qty(-3g - \frac{1}{4g} \mathcal{R}) \,, \\
        I_{\text{fct}} & = \frac{1}{8\pi G_5} \int \dd[4]{x} \sqrt{-\gamma}\qty(\zeta' \mathcal{R}^2 - \zeta'' F_{ij}F^{ij}) \,,
    \end{aligned}
\end{align}
where the integrations are most easily performed over a spacelike hypersurface $ \widehat{r} = \widehat{R}_0$ in Fefferman-Graham coordinates $ (t,\widehat{r},\widehat{\eta},\xi_1,\xi_2)$. The induced metric on this hypersurface is obtained from (\ref{eq-cclp-asymp}). We denote it by $ \gamma_{ij}$ with indices $ i,j = t,\widehat{\eta},\xi_1,\xi_2$ running over the boundary coordinates. The extrinsic curvature and its trace are given, in Fefferman-Graham coordinates, by
\begin{align}
    K_{ij} & = \frac{g \widehat{r}}{2} \pdv{}{\widehat{r}} \gamma_{ij} \,, \quad K = \gamma^{ij} K_{ij} \,.
\end{align}
Finally, $ \mathcal{R}$ is the Ricci scalar calculated from the induced metric $ \gamma_{ij}$. The meaning of these boundary terms is as follows: $ I_{\text{GH}}$ is the usual Gibbons-Hawking boundary term, $ I_{\text{ct}}$ includes the counterterms needed to cancel the divergences of the bulk on-shell action\footnote{Strictly speaking the quoted coefficients of the $ g$ and $ \mathcal{R}$ terms are only valid in Fefferman-Graham coordinates, in other coordinate systems these coefficients might change.}, $ I_{\text{fct}}$ includes the allowed diffeomorphism invariant finite counterterms with arbitrary coefficients $ \zeta'$ and $ \zeta''$. In 5d one also needs to include a non-diffeomorphism invariant conformal anomaly term given by
\begin{align}
    I_{\text{an}} & = \frac{1}{8\pi G_5} \int \dd[4]{x} \sqrt{-\gamma} \, \frac{1}{16g^3}\qty(\mathcal{R}_{ij}\mathcal{R}^{ij} - \frac{1}{3}\mathcal{R}^2 - 4 F_{ij}F^{ij}) \log e^{-2 \widehat{R}_0} \,.
\end{align}
If the integrated conformal anomaly is non-vanishing, generically there is a logarithmic divergence left over that cannot be cancelled. However, on the CCLP solution it is easy to see that, although the un-integrated anomaly does not vanish, its integral does, so $ I_{\text{an}}$ does not contribute at all. Overall, the holographically renormalized Euclidean on-shell action is given by
\begin{align}
    \mathcal{I}_{\text{hr}} & = I_{\mathpzc{E}} + I_{\text{GH}}^{\mathpzc{E}} + I_{\text{ct}}^{\mathpzc{E}} + I_{\text{fct}}^{\mathpzc{E}} \,,
\end{align}
where $ I_{\mathpzc{E}}$ is given by (\ref{eq-action-eucl}) and the superscripts $ \mathpzc{E}$ on the boundary terms signify that we are working with the Euclideanized versions of (\ref{eq-bdy-terms-hren}). Performing the above integrals is arduous and was done with the help of Mathematica. The final answer for the Euclidean on-shell action is given in (\ref{eq-IrenHR}). 

\subsection{Charges: $ (E, \, J_1, \, J_2)$, via holographic renormalization}\label{sec-charges2}
In the Lorentzian description, we can harness holographic renormalization, see \cite{Skenderis:2002wp} for a review, to obtain the energy of the CCLP black hole. First, we construct the renormalized holographic stress energy tensor as, see \cite{Papadimitriou:2005ii} for more details, 
\begin{align}
    T_{ij} & = -\frac{2}{\sqrt{-\gamma}} \fdv{}{\gamma^{ij}}\qty(\mathcal{L}_{\text{GH}} + \mathcal{L}_{\text{ct}} + \mathcal{L}_{\text{fct}}) \nonumber \\
     & = \frac{1}{8\pi G_5} \bigg[K_{ij} - \gamma_{ij}K + 3g\gamma_{ij} - \frac{1}{2g} \mathcal{R}_{ij} + \frac{1}{4g} \mathcal{R} \gamma_{ij} \nonumber\\
     & \qquad \qquad \quad + 4 \zeta'\qty(\mathcal{R}_{ij} \mathcal{R} - \frac{1}{4} \gamma_{ij} \mathcal{R}^2 - \nabla_i \nabla_j \mathcal{R} + \gamma_{ij} \nabla_k \nabla^k \mathcal{R})\bigg] \,,
\end{align}
where we have already set the $ \zeta''$ contribution to zero since $F_{ij}F^{ij}$ vanishes sufficiently fast near the boundary of the CCLP solution. Then, the conserved charge associated to a boundary conformal Killing vector $ K$ can be obtained from 
\begin{align}
    \mathcal{Q}\qty[K] & = \int_{0}^{2\pi}\dd{\xi_1} \int_{0}^{2\pi}\dd{\xi_2} \int_{0}^{\pi/2}\dd{\widehat{\eta}} \sqrt{-\gamma} \, T\indices{^i_j}K^j \,.
\end{align}
In particular, for the two angular momenta we obtain
\begin{align}
    J_1 & = \mathcal{Q}\qty[\partial_{\xi_1}] = \frac{\pi\qty(2am + bq(1 + a^2g^2))}{4 \Xi_a^2 \Xi_b G_5} \,, \quad J_2 = \mathcal{Q}\qty[\partial_{\xi_2}] = \eval{J_1}_{b \leftrightarrow a}
\end{align}
in agreement with our previous Komar integral results (\ref{eq-ang-mom}). For the energy we obtain
\begin{align}
    E = \mathcal{Q}\qty[\partial_t] & = \qty(E_{\text{AdS}} - \widetilde{\zeta}') + \frac{m \pi(2\Xi_a + 2\Xi_b - \Xi_a \Xi_b) + 2abq^2 \pi(\Xi_a + \Xi_b)}{4 \Xi_a^2 \Xi_b^2 G_5} \,,
\end{align}
where $ E_{\text{AdS}}$ and $ \widetilde{\zeta}'$ are the constants given in (\ref{eq-casimir-and-zeta}).

\subsection{Killing spinor in CCLP coordinates}\label{sec-killing-spinor}
Here we show that the three parameter, i.e.~$(a, \, b, \, \widetilde{m})$, CCLP solution, described in Section~\ref{sec-susy-and-bps-limits}, is supersymmetric by explicitly verifying that the following spinor
\begin{align}\label{eq-spinor-susy-cclp-hopf1}
   \epsilon & = \exp\qty{\frac{\iu}{2} \qty[g\qty(1 + \sqrt{3} \mathpzc{x} \alpha)t + \xi_1 + \xi_2]} \frac{1}{\rho}\qty[\widetilde{m} \rho^2  - (1 + \widetilde{m})\qty(r^2 - r_*^2)]^{1/2} \, \epsilon_0 \,,
\end{align}
solves (\ref{eq-bps}). Here $ \epsilon_0$ is a constant spinor that satisfies the projection condition
\begin{align}\label{eq-projector}
    \frac{\iu}{2}\qty(\gamma^{12} - \gamma^{34})\epsilon_0 & = \epsilon_0 \,,
\end{align}
which reduces the number of independent (complex) spinor components from four to one. We use the explicit realization of the Clifford algebra described in \cite{Freedman:2012zz} and the following frames
\begin{align}
    \begin{aligned}
        e^0 & = 
            - \frac{g^2\Delta_\eta}{\Xi_a \Xi_b}\qty[ \frac{(a + b)(1 + \widetilde{m})(1 + g^2 r_*^2)}{g^3 \rho^2} - \qty(g^{-2} + r_*^2)  - \frac{\rho^2}{1 + \widetilde{m}}\qty(1 - \frac{\widetilde{m}\rho}{h})] \dd{t} \\
            & \quad + \frac{g \sin^2\eta}{\Xi_a}\qty[\frac{a(a + b)(1 + \widetilde{m})(1 + g^2 r_*^2)}{g^2 \rho^2} - \qty(a^2 + r_*^2) - \frac{\rho^2}{1 + \widetilde{m}}\qty(1 - \frac{\widetilde{m}\rho}{h})] \dd{\xi_1} \\
            & \quad + \frac{g \cos^2\eta}{\Xi_b}\qty[\frac{b(a + b)(1 + \widetilde{m})(1 + g^2 r_*^2)}{g^2 \rho^2} - \qty(b^2 + r_*^2) - \frac{\rho^2}{1 + \widetilde{m}}\qty(1 - \frac{\widetilde{m}\rho}{h})] \dd{\xi_1} \,, \\
        e^1 & = \sqrt{\frac{\rho^2}{\Delta_r}} \dd{r} \,, \\
        e^2 & = \frac{r \sqrt{\Delta_r \rho^2}}{h} \qty(- \frac{g \Delta_\eta}{\Xi_a \Xi_b} \dd{t} + \frac{\sin^2\eta}{\Xi_a} \dd{\xi_2} + \frac{\cos^2\eta}{\Xi_b} \dd{\xi_1}) \,, \\
        e^3 & = - \sqrt{\frac{\rho^2}{\Delta_\eta}} \dd{\eta} \,,\\
        e^4 & = \frac{\sin\eta \cos\eta \sqrt{\Delta_\eta \rho^2}}{h} \Bigg\{ - \frac{g(\Xi_a - \Xi_b)}{\Xi_a \Xi_b}\qty(g^{-2} + (1 + \widetilde{m})r_*^2 - r^2) \dd{t} \\
        & \qquad \qquad + \frac{1}{\Xi_a}(a^2 + (1 + \widetilde{m})r_*^2 - r^2) \dd{\xi_1} + \frac{1}{\Xi_b}(b^2 + (1 + \widetilde{m})r_*^2 - r^2) \dd{\xi_2}\Bigg\} \,.
    \end{aligned}
\end{align}
In \cite{Cabo-Bizet:2018ehj} a much simpler expressions for the frames in the so called orthotoric coordinates were presented. Our expressions are merely coordinate transformed versions of their (A.4). In these orthotoric coordinates the frame is degenerate in the interesting limits to extremality, $ \widetilde{m} \to 0$, and equal rotation parameters, $ b \to a$. In \cite{Cassani:2015upa} it was explained how to take these limits sensibly by performing accompanying field redefinitions. The benefit of working directly in the CCLP coordinates presented here is that the frame is perfectly valid in both the extremal and the equal rotation parameter limits. One can use the explicit expressions for the spinor and frame to understand the linear constraint (\ref{eq-susy-constraint1}). It arises from evaluating the Lie derivative along the Killing vector $ V$, described in Appendix \ref{sec-charges1}, on $ \epsilon$, at the horizon
\begin{align}
    \mathcal{L}_V \epsilon & = V^\mu \qty(\partial_\mu \epsilon + \omega_{\mu ab}\gamma^{ab}\epsilon) - \frac{1}{4} \nabla_\mu V_\nu \gamma^{\mu\nu} \epsilon \nonumber \\
    & = \frac{\iu}{2}\qty(g + \Omega_1 + \Omega_2 + g \mathpzc{x} \sqrt{3} \, \alpha + 2 \eval{V^\mu A_\mu}_{r = r_+}) \epsilon \,.
\end{align}
We implement the regularity condition $ A^\mu A_\mu < \infty$, by setting $ \alpha = -\Phi$. This removes the $ \eval{V^\mu A_\mu}_{r = r_+}$ term. Now we want to ensure that 
\begin{align}
    \mathcal{L}_V \epsilon & = \frac{\pi}{\beta} \epsilon  \implies \eval{e^{t \mathcal{L}_{V}} \epsilon}_{t = t_0}^{t = t_0 - \iu \beta} = e^{- \iu \beta \mathcal{L}_V} \epsilon = - \epsilon \,,
\end{align}
which implies that the Killing spinor has the correct antiperiodic behavior in the vicinity of the horizon. This is compatible with the Lorentzian signature identification in (\ref{eq-twisted-identification-eucl}). This results in precisely the linear constraint (\ref{eq-susy-constraint1}) between the potentials $ (\beta,\Omega_1,\Omega_2,\Phi)$, where $ \Phi$ appears through $-\alpha$.

\section{Conventions}\label{sec-conventions}
We define forms with explicit factorial factors, for example a $ p$-form $ \omega$ reads
\begin{align}
    \omega & = \frac{1}{p!} \omega_{\mu_1\dots \mu_p} \dd{x^{\mu_1}} \wedge \dots \wedge \dd{x^{\mu_p}} \,.
\end{align}
In particular, we have
\begin{align}
    A & = A_\mu \dd{x^\mu} \,, \quad F = \frac{1}{2!} F_{\mu\nu} \dd{x^\mu} \wedge \dd{x^\nu} \,, \quad F = \dd{A} = \partial_{[\mu} A_{\nu]} \dd{x^\mu} \wedge \dd{x^\nu} \,,
\end{align}
where antisymmetrization is also defined with explicit factorial factors as
\begin{align}
    T_{[\mu_1\dots \mu_p]} & = \frac{1}{p!} \sum_{\sigma \in S_p} \text{sign}(\sigma) T_{\mu_{\sigma(1)}\dots \mu_{\sigma(p)}} \,.
\end{align}
Thus, the components of the field strength are
\begin{align}
    F_{\mu\nu} & = 2! \partial_{[\mu} A_{\nu]} = \partial_\mu A_\nu - \partial_\nu A_\mu \,.
\end{align}
The Hodge star in our conventions is defined to act on a $ p$-form $ \omega$ as
\begin{align}
    \star\, \omega & = \frac{1}{p!(D-p)!}  \omega_{\mu_1\dots \mu_p} \varepsilon\indices{^{\mu_1\dots \mu_p}_{\nu_1\dots \nu_{D-p}}} \dd{x^{\nu_1}} \wedge \dots \wedge \dd{x^{\nu_{D-p}}} \,,
\end{align}
where $ \varepsilon$ is the epsilon pseudo tensor, related to the totally antisymmetric symbol $ \eta$ as 
\begin{align}
    \varepsilon_{\mu_1\dots \mu_D} & = \sqrt{\abs{g}} \, \eta_{\mu_1 \dots \mu_D} \,, \quad \varepsilon^{\mu_1\dots \mu_D} = \frac{\text{sgn}(g)}{\sqrt{\abs{g}}} \eta^{\mu_1 \dots \mu_D} \,.
\end{align} 
We define the Riemann curvature tensor as
\begin{align}
    R\indices{_{\mu\nu}^\rho_\sigma} & = \partial_\mu \Gamma\indices{^\rho_{\nu\sigma}} - \partial_\nu \Gamma\indices{^\rho_{\mu \sigma}} + \Gamma\indices{^\rho_{\mu \lambda}}\Gamma\indices{^\lambda_{\nu\sigma}} - \Gamma\indices{^\rho_{\nu \lambda}}\Gamma\indices{^\lambda_{\mu \sigma}} \,,
\end{align}
and the Ricci tensor and Ricci scalar as
\begin{align}
    R_{\mu\nu} & = R\indices{_\mu^\rho_\nu_\rho} = R\indices{^\rho_{\mu\rho\nu}} \,, \quad R = R\indices{^\mu_\mu} \,.
\end{align}
For the gamma matrices in arbitrary dimensions we use the conventions of \cite{Freedman:2012zz}, in particular, for the $S^4$ and $\Sigma_{\mathfrak{g}}$, described in the main text, we use Euclidean gamma matrices, while for the 7d and the 5d spacetimes we use the Lorentzian ones.


\bibliography{5d_M5_BH}
\bibliographystyle{JHEP}

\end{document}